\documentclass[aps,prx,reprint,groupedaddress,floatfix]{revtex4-1}
\pdfoutput=1

\usepackage{siunitx} % Provide correct typesetting for SI units

\usepackage{amsmath} % Equation typesetting
\usepackage{amssymb}
\usepackage{mathrsfs}
\usepackage{mathtools}
\usepackage{stmaryrd}
\usepackage{color}
\usepackage{esint}

\usepackage[shortlabels]{enumitem}
\usepackage{graphicx} % Image handler

\usepackage{multirow}
\usepackage{booktabs} % Horizontal rules in tables
\usepackage{hyperref} % For hyperlinks in the PDF

\DeclareMathAlphabet{\mathpzc}{OT1}{pzc}{m}{it}

\begin{document}

%----------------------------------------------------------------------------------------
%	TITLE SECTION
%----------------------------------------------------------------------------------------

\title{Orientational wetting and topological transitions in confined solutions of semi-flexible polymers}
\date{\today}
\author{Maxime M. C. Tortora}
\email{maxime.tortora@ens-lyon.fr}
\author{Daniel Jost}
\affiliation{Universit\'e de Lyon, ENS de Lyon, Univ Claude Bernard, CNRS, Laboratoire de Biologie et Mod\'elisation de la Cellule, Lyon, France}

%----------------------------------------------------------------------------------------

%----------------------------------------------------------------------------------------
%	ABSTRACT
%----------------------------------------------------------------------------------------

\begin{abstract}
{\centering
\par
 \includegraphics[height=3.5cm]{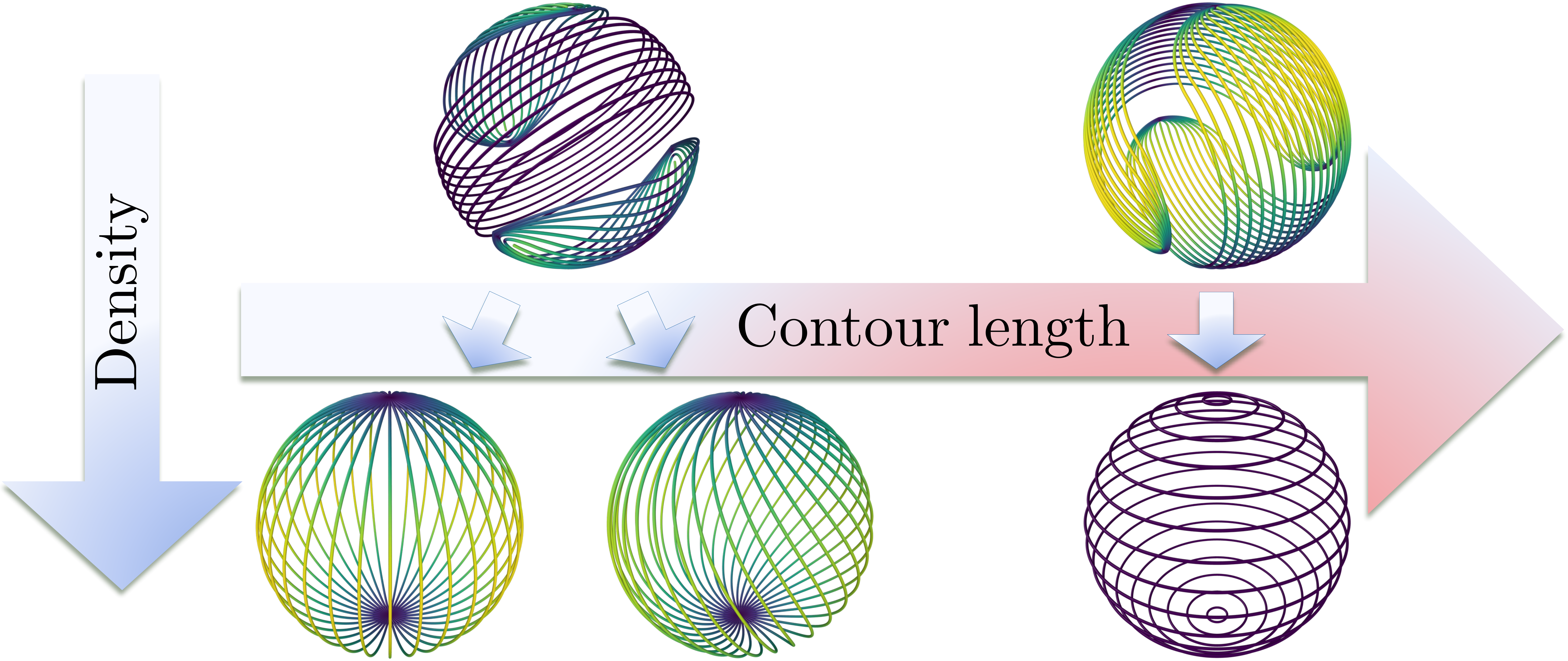}
\par
}

Despite their considerable practical and biological applications, the link between molecular properties, assembly conditions and self-organized structure in confined polymer solutions remains elusive. Here, we explore the lyotropic nematic ordering of semi-flexible chains in spherical confinement for multiple contour lengths across a wide regime of concentrations. We uncover an original crossover from two distinct quadrupolar states, both characterized by regular tetrahedral patterns of surface topological defects, to either longitudinal, latitudinal or spontaneously-twisted bipolar structures with increasing densities. These transitions, along with the intermediary arrangements that they involve, are attributed to the combination of orientational wetting with subtle variations in their liquid-crystal (LC) elastic anisotropies. Our molecular simulations are corroborated by density functional calculations, and are quantified through the introduction of several order parameters as well as an unsupervised learning scheme for the localization of topological defects. Our results agree quantitatively with the predictions of continuum nematic elasticity theories, and evidence the extent to which the folding of macromolecules and the self-assembly of low-molecular-weight LCs may be guided by the same, universal principles.
\end{abstract}

%\pacs{23.23.+x, 56.65.Dy}
%\keywords{Density functional theory, liquid crystals, multi-scale modelling, chirality, perturbation theory.}

\maketitle % Insert title

%----------------------------------------------------------------------------------------
%	ARTICLE CONTENTS
%----------------------------------------------------------------------------------------

\section{Introduction}
The tight packing of macromolecules within confined, crowded domains is commonly observed in a vast array of biological contexts~\cite{ellis2001macromolecular}. Geometrical confinement, which may arise from the presence of physical boundaries such as membrane walls or fluid interfaces, generally reduces the entropy of individual molecules by restricting the ensemble of accessible conformations that they may adopt. These effects are typically supplemented by elastic contributions arising from the finite compliance of intra-molecular bonds, as well as additional steric constraints emerging from the dense bulk environment~\cite{zhou2008macromolecular}. This interplay between entropic and elastic forces is often sufficient to induce liquid-crystalline (LC) behavior at high-enough polymer concentrations, driven by the combination of local shape anisotropy imparted by bending stiffness and excluded-volume contributions resulting from the impenetrability of the macromolecular backbone~\cite{binder2020understanding}. The nematic phase, characterized by the spontaneous alignment of neighboring chain segments, constitutes the simplest instance of such supramolecular structures, and is ubiquitously observed \textit{in vivo} and \textit{in vitro} in dense biopolymer solutions~\cite{hamley2010liquid,mitov2017cholesteric}.
\par
Furthermore, in finite-size systems, the preferred local orientational order of the nematic phase is generally impaired by the global geometry of the confining volume. Within spherical cavities, extensively studied as idealized models for macromolecular packing in biological and microfluidic compartments~\cite{micheletti2011polymers,shaebani2017compaction,curk2019spontaneous}, this frustration commonly manifests through the tangential anchoring of the nematic direction of alignment (known as the \textit{director}) at the sphere surface, which is typically favored by simple steric interactions between polymers and the confining wall or fluid interface. In this case, a celebrated theorem due to Poincar\'e and Hopf~\cite{frankel2003geometry} imposes that nematic organization is disrupted by \textit{topological defects}~\cite{kleman2003topological}, required to yield a total \textit{topological charge} matching the Euler characteristic $s=2$ of the sphere. 
\par
{
The coupling between LC order and geometrical confinement may thus lead to a rich zoo of topological structures, which have spawned considerable experimental and theoretical interest over the last decades~\cite{urbanski2017liquid}. Pioneering work by Lavrentovich, Kl\'eman and others has demonstrated that the defect type, number and morphology may be fine-tuned through the adjustment of boundary conditions, sample size and thickness~\cite{volovik1983topological,lavrentovich1998topological,kleman2006topological,lopez2011frustrated,tomar2012morphological}. Such assemblies have been further evidenced by extensive numerical simulations in a variety of contexts~\cite{chiccoli1990computer,chiccoli1997monte,bradac1998molecular,stark1999water,andrienko2002defect,guzman2003defect,gharbi2013microparticles}, and may be generally accounted for by continuum theories of LC elasticity~\cite{deGennes1993physics}.
\par
However, the vast majority of previous investigations have focused on the \textit{thermotropic} class of LCs, characteristically formed by low-complexity compounds whose molecular sizes fall in the nanometer range. In contrast, studies of confined macromolecular (\textit{lyotropic}) LCs have remained comparatively scarcer, despite notable recent progress~\cite{nikoubashman2017semiflexible,milchev2018densely,khadilkar2018self,nikoubashman2021ordering}, with the majority of theoretical efforts directed towards the limiting case of a single long chain folded within a spherical capsule~\cite{marenduzzo2010biopolymer,shin2011filling,zhang2011tennis,chen2016theory,liang2019orientationally}. Hence, the quantitative link between defect arrangements, polymer mechanics and thermodynamic state remains largely unresolved in such macromolecular systems, whose dimensions may often be comparable to those of the (typically micrometric) confining cavity --- thus casting doubts on the applicability of continuum approaches~\cite{garlea2016finite}.
 \par
Here, we seek to elucidate the interplay between geometrical constraints and LC organization in confined solutions of self-avoiding worm-like chains (SAWLCs) by means of large-scale molecular dynamics and density functional calculations. This coarse-grained model, widely used as a generic template for common persistent (bio)polymers in good solvent conditions~\cite{broedersz2014modeling}, is characterized by purely-repulsive, short-ranged inter-molecular interactions, and therefore undergoes Onsager-like self-ordering transitions driven chiefly by entropy~\cite{binder2020understanding,onsager1949effects} --- as opposed to the enthalpy-dominated phases displayed by thermotropic LCs. Our numerical description and simulation protocol, along with the corresponding theoretical framework, are introduced in detail in Sec.~\ref{sec:materials}. We then outline in Sec.~\ref{sec:results} their application to the nematic phase of DNA-like SAWLCs of diverse lengths confined within discretized spherical shells of continuously-tunable radii. We finally recapitulate our results in Sec.~\ref{sec:discussion}, and highlight their main physical and biological implications.

\section{Materials and methods} \label{sec:materials}

\subsection{Numerical model} \label{sec:model}
Polymers were described through the Kremer-Grest model~\cite{grest1986molecular,kremer1990dynamics}, which corresponds to a discretized realization of the classical Kratky-Porod worm-like chain~\cite{kratky1949diffuse} with excluded volume constraints. In this framework, each macromolecular chain $\mathpzc{C}$ was represented as a linear assembly of $N_m$ monomeric units indexed by $k$, interacting via intra- and inter-molecular steric repulsive forces with effective chain diameter $\sigma$,
\begin{equation}
  \label{eq:poly_exc}
   \mathscr{H}_{\rm poly}^{\rm exc}=\sum_{\mathpzc{C}} \sum_{k,l\in \mathpzc{C}} u^{\rm WCA}_{\sigma}\big(r_{kl}\big) + \sum_{\mathpzc{C},\mathpzc{C}'}\sum_{\substack{k\in\mathpzc{C} \\ k' \in\mathpzc{C}'}} u^{\rm WCA}_{\sigma}\big(r_{kk'}\big),
\end{equation}
in which comma-separated indices imply summation over unique pairs of distinct elements and $u^{\rm WCA}$ is the Weeks-Chandler-Andersen (WCA) potential~\cite{weeks1971role},
\begin{equation*}
  u^{\rm WCA}_{\sigma}(r)= 
  \begin{dcases}
    4\epsilon \Bigg [ \bigg(\frac{\sigma}{r} \bigg)^{12} - \bigg( \frac{\sigma}{r}\bigg)^6 + \frac{1}{4} \Bigg] & \text{\!if } r < 2^{1/6}\sigma \\
    0 &\text{\!if } r \geq 2^{1/6}\sigma
  \end{dcases},
\end{equation*}
where $\epsilon$ defines the model unit of energy. Chain connectivity was enforced by linking adjacent monomers via finitely-extensible non-linear elastic (FENE) springs,
\begin{equation*}
  u^{\rm FENE}_{\rm poly}\big(\Delta_k\big)= 
  \begin{dcases}
    -\frac{k_c r_c^2}{2} \log\Bigg[1-\bigg(\frac{\Delta_k}{r_c}\bigg)^2\Bigg] & \text{\!if } \Delta_k < r_c \\
    0 &\text{\!if } \Delta_k \geq r_c
  \end{dcases},
\end{equation*}
in which $\Delta_k\equiv \lVert \mathbf{r}_k-\mathbf{r}_{k-1}\rVert$ is the corresponding bond length, with $\mathbf{r}_k$ the center-of-mass position of the $k$-th bead. Chain stiffness is then governed by a simple angular potential of the form~\cite{grest1986molecular,kremer1990dynamics}
\begin{equation*}
  u^{\rm bend}_{\rm poly}(\Theta_k) = \kappa_b \big(1-\cos\Theta_k\big),
\end{equation*}
where $\Theta_k \equiv \arccos\big(\widehat{\mathbf{t}}_k\cdot\widehat{\mathbf{t}}_{k+1}\big)$ is the inter-bond angle associated with the triplet of consecutive monomers $\big(k-1,k,k+1\big)$, with $\widehat{\mathbf{t}}_k\equiv \big(\mathbf{r}_{k}-\mathbf{r}_{k-1}\big)/\Delta_k$ the bond unit vector. We set $r_c=1.5\,\sigma$ and $k_c=30\,\epsilon/\sigma^2$, so that the individual bond length resulting from the competition between WCA repulsion and FENE stretching remains effectively constant throughout the simulations, and is given by $\Delta_k =  l_b \simeq 0.97\,\sigma$~\cite{egorov2016insight}. In this case, the polymer persistence length is simply related to the bending modulus $\kappa_b$ via $l_p\simeq \beta \kappa_b l_b$, with $\beta \equiv 1/k_BT$ the inverse temperature, which holds for stiff chains ($\beta \kappa_b \gtrsim 2$)~\cite{hsu2010standard}. Since we here restrict our focus to the case of the nematic phase, we set the polymer persistence length to $l_p=25\,\sigma$ to emulate the relative bending rigidity of DNA duplexes ($\sigma \sim \SI{2}{nm}$)~\cite{broedersz2014modeling}, which generally do not display significant stable smectic order in bulk~\cite{salamonczyk2016smectic}. A fixed number $N=\num[group-separator={,},group-minimum-digits={3}]{32768}$ of monomers was used in all simulations, which were evenly distributed among a number $N_c = N/N_m$ of identical chains with contour length $l_c=\big(N_m-1\big)\,l_b+\sigma\simeq N_m\, \sigma$ and volume $v_c\simeq \pi l_c \sigma^2/4$.
\par
The confining membrane was represented as an amorphous polymerized shell, obtained by randomly and uniformly distributing a number $N_v$ of vertices on the surface of a sphere of initial radius $R_0$. The membrane topology was constructed by computing the Delaunay triangulation of the vertex positions via the QuickHull algorithm~\cite{barber1996quickhull}. Each generated bond was described by a harmonic potential of stiffness $k_m$, which yields a total stretching energy~\cite{kantor1987phase}
\begin{equation*}
  \mathscr{H}_{\rm memb} = \frac{k_m}{2} \sum_{i,j} \Big(r_{ij} - r_{ij}^0\Big)^2,
\end{equation*}
where the sum runs over all pairs of linked vertices $i$ and $j$, denoting by $r_{ij}^0\simeq r_0 = \big(8\pi R_0^2/\sqrt{3}N_v\big)^{0.5}$ their separation distance in the relaxed spherical reference state. The coupling between macromolecules and confining membrane was introduced through a steric repulsion term involving each heterogeneous pair of vertices $i\in\llbracket1,N_v\rrbracket$ and monomers $k$,
\begin{equation*}
  \mathscr{H}^{\rm poly}_{\rm memb} = \sum_{i=1}^{N_v}\Bigg(\sum_{\mathpzc{C}} \sum_{k\in \mathpzc{C}} u^{\rm WCA}_{\Sigma}\big(r_{ik}\big)\Bigg),
\end{equation*}
where the effective excluded diameter $\Sigma = \big(r_0+\sigma\big)/2$ was chosen based on Lorentz-Berthelot combining rules to ensure that the shell surface may not be crossed by the encapsulated chains in any conformation. Although membrane mechanics were not explicitly investigated here, we set $k_m = 20 \, k_BT/\sigma^2$, which yields a bare Young's modulus $Y_0=2k_m/\sqrt{3}$ in the continuum limit~\cite{kantor1987phase} approaching the value $Y_0 \simeq\SI{25}{\milli\newton\per\meter}$ measured for the nuclear envelope lamina network at temperature $T=\SI{300}{\kelvin}$~\cite{dahl2004nuclear}. A total number $N_v=\num[group-separator={,},group-minimum-digits={3}]{9800}$ of vertices was used in all simulations. The complete Hamiltonian then reads as $\mathscr{H}=\mathscr{H}_{\rm memb}+\mathscr{H}_{\rm poly}+\mathscr{H}^{\rm poly}_{\rm memb}$, with $\mathscr{H}_{\rm poly}=\mathscr{H}_{\rm poly}^{\rm FENE}+\mathscr{H}_{\rm poly}^{\rm bend}+\mathscr{H}_{\rm poly}^{\rm exc}$, which fully specifies the mechanics of the polymer-membrane system. 
\par
Molecular dynamics simulations were conducted in the canonical ensemble as realized by a Langevin thermostat at fixed temperature $T=\epsilon/k_B$, and were evolved via a standard velocity-Verlet integration scheme~\cite{frenkel2002understanding}. Calculations were performed on multiple graphics processing units (GPUs) using the HOOMD-blue software package~\cite{glaser2015strong,anderson2020hoomd}. Polymer chains were initially arranged in a simple cubic crystalline configuration, and the initial membrane radius was set to $R_0=4\,l_c$ to ensure that the starting state of all simulations lied deep in the stability range of the disordered (isotropic) phase and in conditions of weak confinement. Relaxation runs of $\mathcal{O}\big(10^9\big)$ MD steps were performed to achieve full decorrelation from the initial orientational and positional order. Production runs then consisted in a slow isothermal compression carried out by gradually reducing the reference length $r_0$ of the membrane springs in $\mathcal{O}\big(10^2\big)$ increments. A stationary state was typically reached in $\mathcal{O}\big(10^8\big)$ MD steps following each membrane volume move, after which measurements were performed over a further $\mathcal{O}\big(10^8\big)$ MD steps. Equilibration was evaluated by comparing the free energy obtained from simulations via the compression route against the predictions of density functional theory (DFT) for the same polymer model (see Sec.~\ref{sec:dft}). However, convergence towards long-lived, metastable states cannot be entirely ruled out in such hysteresis-prone systems of dense, confined polymers, and could potentially manifest --- especially in the vicinity of the isotropic-nematic (I-N) transition region~\cite{nikoubashman2017semiflexible,milchev2018densely}. The density regimes thus explored by our simulations span nearly 3 orders of magnitudes for the longest chains studied ($l_c=64\,\sigma$), which was made computationally feasible by the use of state-of-the-art tree-based neighbor list calculations~\cite{howard2016efficient}. The corresponding polymer volume fractions $\eta$ were computed as 
\begin{equation*}
  \eta\equiv N_c v_c/V, 
\end{equation*}
where the membrane volume $V$ was evaluated using standard algorithms for closed and oriented triangular surface meshes~\cite{zhang2001efficient}. The mean radius $R$ of the thermalized shell subsequently reads as $R\equiv (3V/4\pi)^{1/3}$, and covers an approximate range $R\in [25\,\sigma, 256\,\sigma]$, while the initial polymer packing fractions respectively read as $\eta_0\simeq2\,\%$ ($l_c=16\,\sigma$), $\eta_0\simeq0.3\,\%$ ($l_c=32\,\sigma$) and $\eta_0\simeq0.04\,\%$ ($l_c=64\,\sigma$) (see Fig.~S1).

\subsection{Describing non-uniform orientational order} \label{appendix:describing}

To quantify nematic organization in assemblies of apolar macromolecules, it is customary to introduce the traceless Landau-de Gennes $\mathcal{Q}$-tensor~\cite{deGennes1993physics},
\begin{equation}
  \label{eq:deGennes}
  \mathcal{Q}_k^{\alpha\beta} = \frac{3\,\widehat{t}_k^\alpha \widehat{t}_k^\beta - \delta_{\alpha\beta}}{2},
\end{equation}
where $\alpha,\beta\in \{x,y,z\}$ denote the components of a bond vector $\widehat{\mathbf{t}}_k$ in the fixed laboratory frame and $\delta$ is the Kronecker delta. In the following, we omit the hat notation from unit vectors when no confusion can arise. In spatially-uniform systems at equilibrium, one may probe the potential collective anisotropy of the structure from the ascending eigenvalues $\lambda_1\leq \lambda_2 \leq \lambda_3$ of the tensor $\mathcal{Q}\equiv\big\langle\mathcal{Q}_k\big\rangle$, with $\big\langle\cdot\big\rangle$ a time and ensemble average over all constituent bonds $k$. In this case, the eigenvector $\mathbf{e}_3$ associated with $\lambda_3$ may be identified as the preferred direction of alignment $\mathbf{n}_0$ of the chains, referred to as the \textit{nematic director}, while $\lambda_3$ is related to the angular dispersion $\theta_k \equiv \arccos \big(\mathbf{t}_k \cdot \mathbf{n}_0\big)$ of bonds $\mathbf{t}_k$,
\begin{equation}
  \label{eq:legendre}
  \lambda_3 = \frac{3\big\langle\cos^2 \theta_k\big\rangle-1}{2}\equiv \big\langle P_2\big(\cos\theta_k\big)\big\rangle,
\end{equation}
with $P_2$ the second Legendre polynomial.
\par
However, in confined systems, the loss of homogeneity imposed by the presence of physical boundaries implies that both the director and the local degree of orientational order are generally position-dependent. Furthermore, the geometry of the confining walls may induce additional breakings of local phase symmetries. For instance, in the isotropic phase, spherical invariance dictates $\lambda_1=\lambda_2=\lambda_3\,(=0)$ in the bulk, while the inhibition of chain fluctuations along one direction --- which may arise (e.g.) from the vicinity of a planar interface --- typically yields $\lambda_1 < \lambda_2=\lambda_3\,(=-\lambda_1/2)$ near a flat boundary, where the eigenvector $\mathbf{e}_1$ is borne by the plane normal. Conversely, one expects $\lambda_3 > \lambda_1=\lambda_2\,(=-\lambda_3/2)$ in the bulk nematic phase of uniaxial molecules such as SAWLCs, reflecting the cylindrical symmetry of the equilibrium chain distribution about $\mathbf{e}_3=\mathbf{n}_0$ --- while the existence of a preferred direction of alignment lifts the degeneracy in $\lambda_2$ and $\lambda_3$ at the confining walls, thus leading to a biaxial surface structure ($\lambda_1 < \lambda_2< \lambda_3$).
\par
Hence, the interpretation of such tensorial order parameters (OPs) is necessarily ambiguous in finite-size samples. At the local level, it nonetheless follows from the previous discussion that a generic alignment parameter $\alpha \in [0,1]$ may be introduced to measure the degree of nematic order,
\begin{equation}
  \label{eq:alpha}
  \alpha \equiv \frac{2\big(\lambda_3-\lambda_2\big)}{3}.
\end{equation}
Eq.~\eqref{eq:alpha} satisfies $\alpha=0$ in the absence of local alignment ($\lambda_2=\lambda_3$) and $\alpha=1$ in the limit of perfect orientational order ($\lambda_3=1$, $\lambda_2=-1/2$), both in the bulk and near the boundary, and simply reduces to $\alpha=\lambda_3$ in the case where the order is purely uniaxial ($\lambda_2=-\lambda_3/2)$. 
\par
To unequivocally quantify orientational order in confined, inhomogeneous LCs, it is thus necessary to devise a hierarchy of OPs capable of discerning the \textit{global} symmetries of the system, as described by the long-ranged spatial patterns of the director field $\mathbf{n}$, from the \textit{local} structure arising from the detailed arrangements of the polymers about $\mathbf{n}$. For this purpose, we discretize the volume $V$ enclosed by the membrane into $\mathcal{O}\big(10^2\big)$ spherical volume elements $\Xi_{\mathbf{r}}$, where $\mathbf{r} \equiv (r,\theta,\phi)$ denotes spherical coordinates, and introduce the locally-averaged tensor
\begin{equation}
  \label{eq:q_loc}
  \mathcal{Q}^{\alpha\beta}(\mathbf{r}) = \Big\langle\mathcal{Q}_k^{\alpha\beta}\Big\rangle_{\Xi_{\mathbf{r}}},
\end{equation}
in which $\big\langle\cdot\big\rangle_{\Xi_{\mathbf{r}}}$ is an ensemble average over all bonds $k$ whose center of mass lies within $\Xi_{\mathbf{r}}$. We may then quantify local nematic order at position $\mathbf{r}$ through the alignment parameter $\alpha(\mathbf{r})$ associated with the eigenvalues $\lambda_i(\mathbf{r})$ of $\mathcal{Q}(\mathbf{r})$ (Eq.~\eqref{eq:alpha}), and define the spatially-resolved director field $\mathbf{n}(\mathbf{r})\equiv \mathbf{e}_3(\mathbf{r})$ as the eigenvector associated with $\lambda_3(\mathbf{r})$. Note that to improve statistical sampling, Eq.~\eqref{eq:q_loc} is further averaged over multiple equilibrated configurations prior to spectral analysis~\cite{trukhina2008computer}. We may finally describe long-wavelength fluctuations of the director field over any collection $\Omega$ of volume elements $\Xi_{\mathbf{r}}$ via the eigenvalues $\Lambda_i(\Omega)$ of the tensor
\begin{equation}
  \label{eq:q_dir}
  \mathcal{S}^{\alpha\beta}(\Omega) = \int_{\mathbf{r}\in\Omega} d^3\mathbf{r} \, w_\Omega(\mathbf{r}) \frac{3\,n^\alpha(\mathbf{r})n^\beta(\mathbf{r})-\delta_{\alpha\beta}}{2},
\end{equation}
where the weight $w_\Omega(\mathbf{r}) \equiv \alpha(\mathbf{r}) / \int_\Omega\alpha$ effectively subdues contributions to the integral Eq.~\eqref{eq:q_dir} from regions of low orientational order, in which the local director $\mathbf{n}(\mathbf{r})$ and its vector components $n^{\alpha,\beta}$ are ill-defined. We here define $\Omega$ as the outer shell of the sample, $\Omega = \big\{\Xi_{\mathbf{r}} \mid r > R-r_s\big\}$, where the thickness of the shell is set to $r_s\equiv 0.15\,R$~\cite{nikoubashman2017semiflexible,milchev2018densely}. The implications of this arbitrary choice are further discussed in Sec.~\ref{sec:quantifying}.

\subsection{Density functional theory of LC elasticities} \label{sec:dft}

In the continuum limit, the free energy of a confined nematic system may be written in the general form~\cite{deGennes1993physics}
\begin{equation}
  \label{eq:f_tot}
   \mathscr{F} = \mathscr{F}_0 + \int_V d\mathbf{r}\, \mathpzc{f}_d(\mathbf{r}),
\end{equation}
where $\mathscr{F}_0$ is the free energy of a nematic phase with uniform director $\mathbf{n}_0$, and $\mathpzc{f}_d(\mathbf{r})$ is the elastic free energy density associated with distortions of the director $\mathbf{n}(\mathbf{r})$~\cite{deGennes1993physics},
\begin{multline}
  \label{eq:f_frank}
  \mathpzc{f}_d= \frac{K_{11}}{2} \big(\nabla\cdot\mathbf{n}\big)^2 + \frac{K_{22}}{2} \big(\mathbf{n} \cdot [\nabla\times\mathbf{n}]\big)^2+ \frac{K_{33}}{2} \big(\mathbf{n}\times[\nabla\times\mathbf{n}]\big)^2 \\ - K_{24} \big(\nabla \cdot \big[\mathbf{n}\{\nabla\cdot \mathbf{n}\} + \mathbf{n}\times\{\nabla\times \mathbf{n}\} \big]\big).
\end{multline}
\par
In Eq.~\eqref{eq:f_frank}, $K_{11}$, $K_{22}$ and $K_{33}$ --- collectively referred to as the Oseen-Frank (OF) elastic moduli --- respectively quantify the thermodynamic penalties incurred by splay, twist and bend deformations, and generally depend on both the molecular system considered and the local degree $\alpha(\mathbf{r})$ of orientational alignment~\cite{deGennes1993physics}. The $K_{24}$ term, known as the \textit{saddle-splay} elasticity, takes the form of a total divergence, and therefore integrates out to a boundary contribution in the limit of homogeneous orientational order ($\alpha(\mathbf{r})=\alpha$), which is typically neglected in studies of bulk phases. This approximation is, however, generally inappropriate in confined geometries~\cite{kos2016relevance}, implying that $K_{24}$ may not be discarded \textit{a priori} in our case. Note that due to the strong tangential anchoring of the chains ($\overline{P}_2(r\to R)\simeq -1/2$, Figs.~\ref{fig1}a-c, inset), we do not include an explicit surface coupling term in Eq.~\eqref{eq:f_frank}, which may be regarded as an external constraint for the purpose of variational analysis as long as the membrane does not significantly deviate from a spherical conformation~\cite{williams1986two}. This strong tangential alignment is consistent with the high anchoring energies reported in nematic melts of SAWLCs, as evaluated in the case of planar walls based on a density functional approach conceptually similar to that outlined below~\cite{terentjev1995density}.

\begin{figure*}[htpb]
  \includegraphics[width=1.5\columnwidth]{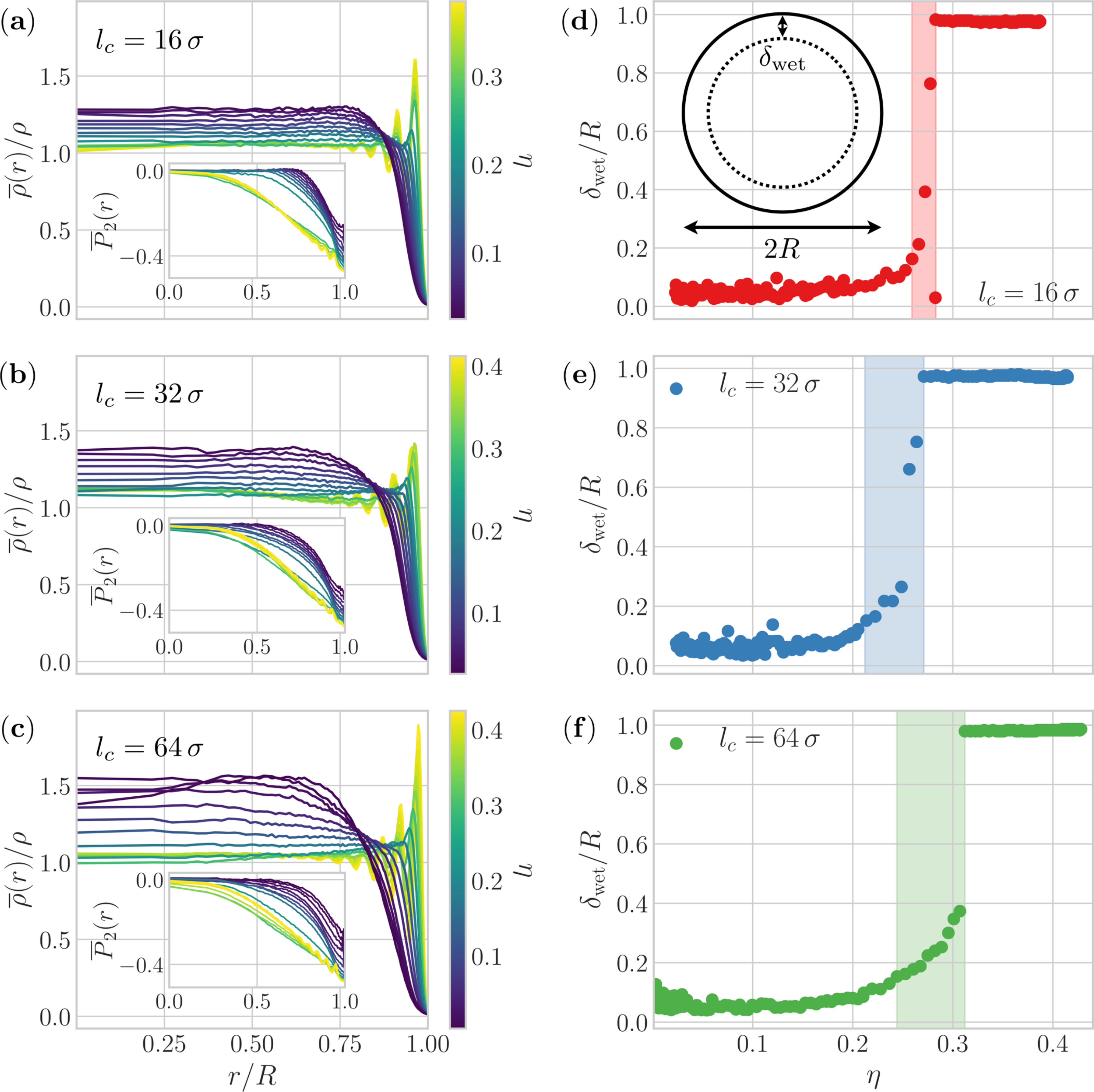}
  \caption{\label{fig1}Orientational wetting behavior of confined DNA-like chains ($l_p=25\,\sigma$). (a)-(c) Radial monomer density $\overline{\rho}(r)\equiv \big\langle \rho(\mathbf{r}) \big\rangle_r$ for various polymer volume fractions $\eta\equiv \rho v_c/N_m$ and contour lengths $l_c$, where $\big\langle\cdot\big\rangle_r$ denotes a volume average over a thin shell of radius $r$. Inset: Mean projection $\overline{P}_2(r) \equiv \big\langle P_2\big(\mathbf{t}_k \cdot \widehat{\mathbf{r}}_k\big)\big\rangle_r$ of bond vectors $\mathbf{t}_k$ on the unit radial vector $\widehat{\mathbf{r}}_k$, with $P_2$ the second Legendre polynomial (Eq.~\eqref{eq:legendre}). $\overline{P}_2(r)=-1/2$ signifies that all chain segments at radial distance $r$ lie in the local tangent plane of the membrane ($\mathbf{t}_k \cdot \widehat{\mathbf{r}}_k=0$), while $\overline{P}_2(r)=0$ indicates that chain orientations are uncorrelated with the membrane surface. (d)-(f) Density variations of the thickness $\delta	_{\rm wet}\equiv R-r_{\rm min}$ of the nematic wetting layer, where $r_{\rm min}$ is defined as the radius of the innermost shell such that $\alpha(r_{\rm min}) \equiv \big\langle\alpha(\mathbf{r})\big\rangle_{r=r_{\rm min}} > 0.5$ (Eq.~\eqref{eq:alpha}). Colored regions mark the approximate location of the partial orientational wetting regime for the different values of $l_c$.}
\end{figure*}

To proceed, we remark that the alignment parameter $\alpha(\mathbf{r})$ does not appreciably depend on position within the nematic regions of the system, outside the microscopic core of the defects (see Sec.~\ref{sec:results}) --- although some moderate layering is apparent at higher densities near the membrane surface (Figs.~\ref{fig1}a-c). Hence, $\alpha(\mathbf{r})\equiv\alpha$ may be taken as uniform throughout the nematic phase as a first approximation. In this case, the saddle-splay contribution reduces to an irrelevant additive constant~\cite{williams1986two}, and the homogeneous OF moduli $K_{ii}=K_{ii}(\alpha)$ may be conveniently calculated by treating Eq.~\eqref{eq:f_tot} as a perturbative expansion with respect to the uniform and uniaxial reference nematic state with director $\mathbf{n}_0$~\cite{tortora2017perturbative}. A microscopic expression for the corresponding reference free energy $\mathscr{F}_0$ may be obtained in the generic functional form~\cite{woodward1991density}
\begin{equation}
  \label{eq:f}
  \mathscr{F}_0[\psi] = \mathscr{F}_{\rm id}[\psi] + \mathscr{F}_{\rm ex}[\psi],
\end{equation}
with $\psi \equiv \psi\big(\mathbf{u}\cdot\mathbf{n}_0\big)$ the molecular distribution quantifying the probability of finding a chain with arbitrary orientation $\mathbf{u}$ at any point in space. The \textit{ideal} component $\mathscr{F}_{\rm id}$ of the free energy then simply corresponds to the entropy of an anisotropic ideal gas of macromolecules~\cite{woodward1991density},
\begin{multline}
  \label{eq:f_id}
  \frac{\beta \mathscr{F}_{\rm id}[\psi]}{V} = \rho_m(\log\rho_m-1)\\+\rho_m\oint d\mathbf{u}\,\log\Big\{\psi\big(\mathbf{u}\cdot\mathbf{n}_0\big)\Big\}\psi\big(\mathbf{u}\cdot\mathbf{n}_0\big),
\end{multline}
with $\rho_m \equiv N_c/V=\eta/v_c$ the molecular density. 
\par
Conversely, the exact functional form of the \textit{excess} contribution $\mathscr{F}_{\rm ex}$, which accounts for the presence of inter-molecular interactions, is generally not known. A successful ansatz due to Fynewever and Yethiraj~\cite{fynewever1998phase} yields $\mathscr{F}_{\rm ex}$ as a direct extension of the Onsager second-virial expression~\cite{onsager1949effects},
\begin{multline}
  \label{eq:f_exc}
  \frac{\beta \mathscr{F}_{\rm ex}[\psi]}{V} = -\frac{\rho_m^2 G(\eta)}{2} \int_V d\mathbf{r} \oiint d\mathbf{u}_1d\mathbf{u}_2 \\ \times \overline{f}\big(\mathbf{r},\mathbf{u}_1,\mathbf{u}_2\big) \psi\big(\mathbf{u}_1\cdot \mathbf{n}_0\big) \psi\big(\mathbf{u}_2\cdot \mathbf{n}_0\big),
\end{multline}
where the prefactor $G(\eta)=(1-0.75\,\eta)/(1-\eta)^2$ is a rescaling correction to approximate the effects of higher-order virial terms based on the Carnahan-Starling equation of state~\cite{parsons1979nematic,lee1987numerical}. In Eq.~\eqref{eq:f_exc}, $\overline{f}\equiv \big\langle\big\langle f_{\mathpzc{C}_1\mathpzc{C}_2}\big\rangle\big\rangle$ is the conformational average of the Mayer $f$-function, which involves the inter-molecular component of Eq.~\eqref{eq:poly_exc} for two arbitrary chain conformations $\mathpzc{C}_{1,2}$ with center-of-mass separation $\mathbf{r}$ and respective orientations $\mathbf{u}_{1,2}$~\cite{tortora2018incorporating},
\begin{equation}
  \label{eq:mayer}
  f_{\mathpzc{C}_1\mathpzc{C}_2}(\mathbf{r},\mathbf{u}_1,\mathbf{u}_2)=\exp\Bigg\{-\beta\sum_{k\in\mathpzc{C}_1} \sum_{k'\in\mathpzc{C}_2} u^{\rm WCA}_{\sigma}\big(r_{kk'}\big)\Bigg\}-1.
\end{equation}
\par
At thermodynamic equilibrium, the most probable distribution $\psi_\eta^{\rm eq}$ proceeds from the functional minimization of Eqs.~\eqref{eq:f}--\eqref{eq:f_exc} at fixed density $\eta$, and may be obtained using standard numerical methods detailed elsewhere~\cite{tortora2018incorporating}. Note that unlike previous investigations, which had to resort to approximate empirical expressions for the generalized excluded volume $\overline{f}$ in Eq.~\eqref{eq:f_exc}~\cite{egorov2016insight,milchev2018nematic,binder2020understanding}, we here evaluate $\overline{f}$ directly from Eq.~\eqref{eq:mayer} using representative ensembles of polymer conformations generated from our simulations. Hence, we are able to accurately apply this density functional theory (DFT) framework to the exact molecular model employed in our simulations~\cite{tortora2017hierarchical,tortora2018incorporating}. 
\par
Let us set the $z$-axis of the laboratory frame to the homogeneous director $\mathbf{n}_0$. Using Eqs.~\eqref{eq:f_tot}--\eqref{eq:f_exc}, lengthy but straightforward manipulations~\cite{tortora2020chiral} yield the OF elastic moduli in the form~\cite{tortora2017perturbative}
\begin{equation}
\label{eq:K1}
\begin{split}
 \!\!\!\!\beta K_{11}^{\rm bare} &= \frac{\rho^2_m G(\eta)}{2} \int_V d\mathbf{r} \oiint d\mathbf{u}_1 d\mathbf{u}_2 \\&\quad\times \overline{f}\big(\mathbf{r},\mathbf{u}_1,\mathbf{u}_2\big) \dot{\psi}_\eta^{\rm eq}(u_{1z}) \dot{\psi}_\eta^{\rm eq}(u_{2z}) r_x^2 u_{1x} u_{2x},
\end{split}
\end{equation}
\begin{equation}
  \label{eq:K2}
  \begin{split}
  \beta K_{22} &= \frac{\rho^2_m G(\eta)}{2} \int_V d\mathbf{r} \oiint d\mathbf{u}_1 d\mathbf{u}_2 \\&\quad\times \overline{f}\big(\mathbf{r},\mathbf{u}_1,\mathbf{u}_2\big) \dot{\psi}_\eta^{\rm eq}(u_{1z}) \dot{\psi}_\eta^{\rm eq}(u_{2z}) r_x^2 u_{1y} u_{2y},
  \end{split}
  \end{equation}
  \begin{equation}
  \begin{split}
  \label{eq:K3}
  \beta K_{33} &=\frac{\rho^2_m G(\eta)}{2} \int_V d\mathbf{r} \oiint d\mathbf{u}_1 d\mathbf{u}_2 \\&\quad\times \overline{f}\big(\mathbf{r},\mathbf{u}_1,\mathbf{u}_2\big) \dot{\psi}_\eta^{\rm eq}(u_{1z}) \dot{\psi}_\eta^{\rm eq}(u_{2z}) r_z^2 u_{1x} u_{2x},
  \end{split}
  \end{equation}
where $\dot{\psi}$ denotes the first derivative of $\psi$. Note that Eqs.~\eqref{eq:K1}--\eqref{eq:K3} rely on the so-called \textit{quasi-homogeneous approximation}~\cite{yokoyama1997density}, which postulates that the local molecular density $\rho_m$ remains unaffected by orientational fluctuations. This assumption is generally inadequate in the case of long macromolecules, for which splay deformations necessarily incur an additional entropic penalty due to the local accumulation of chain extremities~\cite{meyer1982macroscopic}. This effect gives rise to an effective renormalization of the splay modulus $K_{11} = K_{11}^{\rm bare}+\Delta K_{11}$, where the compressibility correction $\Delta K_{11}$ may be approximated as~\cite{meyer1982macroscopic,milchev2018nematic}
\begin{equation}
  \label{eq:delta_K1}
  \beta \Delta K_{11} = \frac{4\eta}{\pi} \frac{l_c}{\sigma^2}.
\end{equation}
The theoretical nematic OP $\alpha=\lambda_3$ at given density $\eta$ finally reads as
\begin{equation}
  \label{eq:alpha_th}
  \alpha = \alpha_{\rm mol} \times \alpha_{\rm bond},
\end{equation}  
where $\alpha_{\rm mol}$ and $\alpha_{\rm bond}$ respectively quantify the alignment of macromolecular axes $\mathbf{u}$ and internal bonds $\mathbf{t}_k$~\cite{tortora2018incorporating},
\begin{equation*}
\begin{gathered}
  \alpha_{\rm mol} = \oint d\mathbf{u} \, \psi_\eta^{\rm eq}\big(\mathbf{u}\cdot\mathbf{n}_0\big) \times P_2\big(\mathbf{u}\cdot \mathbf{n}_0\big), \\
  \alpha_{\rm bond} = \big\langle P_2\big(\mathbf{t}_k\cdot \mathbf{u}\big)\big\rangle.
  \end{gathered}
\end{equation*}
\par
To assess the validity of the theory, we report in Fig.~S2 the density variations of the equilibrium free energy evaluated from Eqs.~\eqref{eq:f}--\eqref{eq:mayer}, compared against that computed by thermodynamic integration along the compression path of our simulations (see SI). The good quantitative agreement obtained over most of the phase diagram for all systems considered suggests that the local structure of the nematic phase is remarkably unaltered by the effects of confinement, thus validating the premise of the perturbative treatment of director distortions underlying Eqs.~\eqref{eq:K1}--\eqref{eq:K3}~\cite{tortora2017perturbative}. Limited underestimations of the free energy are nonetheless apparent at higher densities for short chains with $l_c=16\,\sigma$ (Fig.~S2a), which may reflect potential shortcomings of the rescaled virial approximation (Eq.~\eqref{eq:f_exc})~\cite{egorov2016insight}.

\section{Results} \label{sec:results}

\subsection{From quadrupolar to bipolar order}

\begin{figure*}[htpb]
  \includegraphics[width=2\columnwidth]{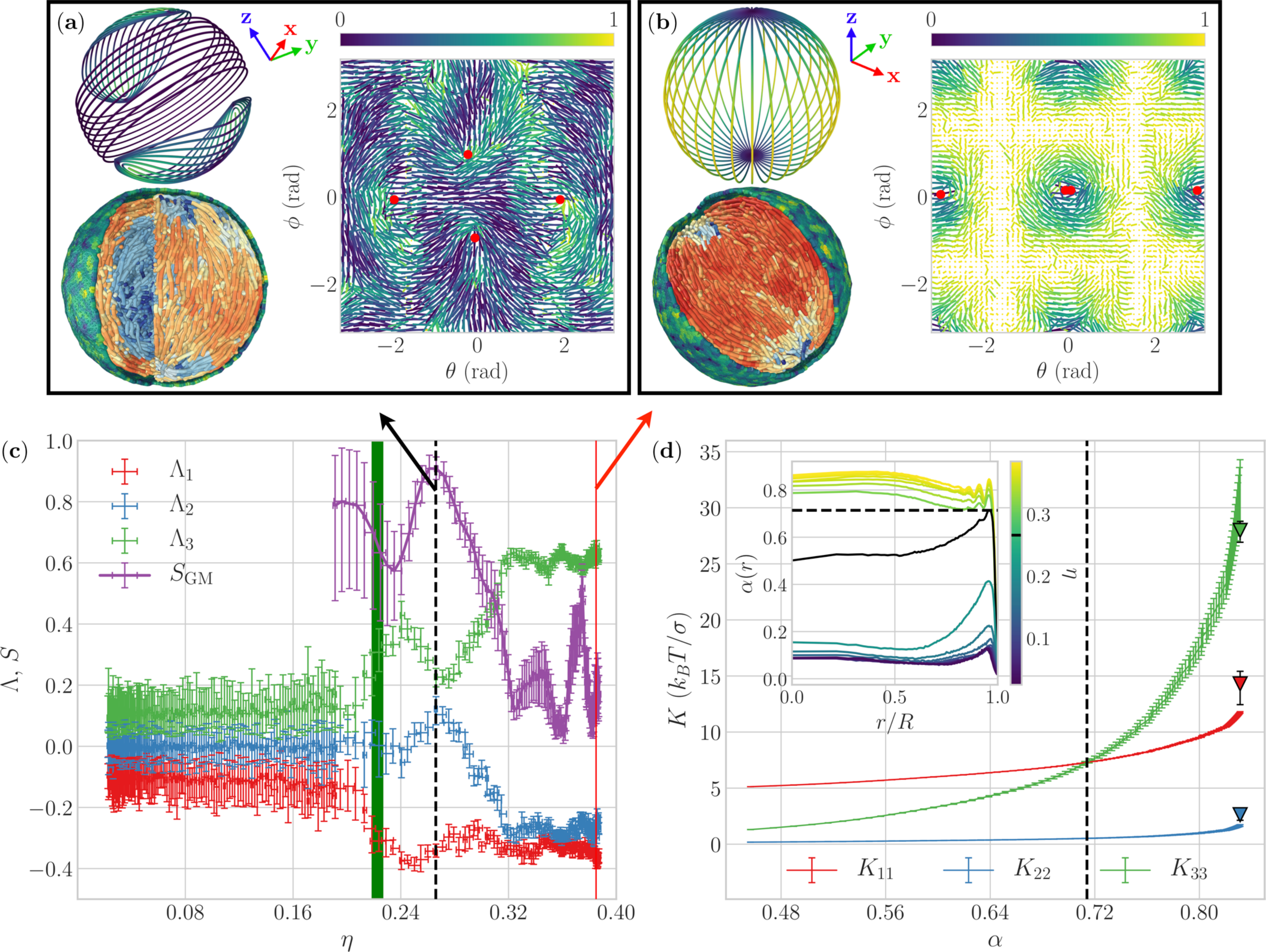}
  \caption{\label{fig2}Self-organization of short DNA-like chains ($l_c=16\,\sigma$) in spherical confinement. (a) Top: Sketch of the CB surface texture. The axis tripod represents the eigenvectors of $\mathcal{S}$ (Eq.~\eqref{eq:q_dir}), with eigenvalues $\Lambda_x = \Lambda_y=-\Lambda_z/2 > 0$ (Table~\ref{table1}). Bottom: Cut and peeled simulation snapshot at volume fraction $\eta\simeq 0.26$. Chains are colored according to $\alpha(\mathbf{r})$, with red (blue) regions corresponding to strong (weak) local alignment. Right: Stereographic map of the surface director field~\cite{vitelli2006nematic} as obtained from simulations. $\phi$ and $\theta$ are the azimuthal and polar angles relative to the $\mathbf{x}$ and $\mathbf{y}$ axes, respectively. Headless arrows are director projections onto the $x$-$y$ plane at position $(\theta,\phi)$, and colors denote the $z$-component of the director. Red dots mark the computed positions of topological defects. (b) Same as (a) for the longitudinal bipolar configuration ($\eta\simeq 0.39$), with $\Lambda_x=\Lambda_y=-\Lambda_z/2 < 0$. (c) Surface OPs ($\Lambda_i$, Eq.~\eqref{eq:q_dir}) and defect Glassmeier parameter ($S_{\rm GM}$, Eq.~\eqref{eq:glassmeier}) as a function of $\eta$. The I-N bulk coexistence region calculated by DFT is highlighted in green. Error bars are standard deviations from 2 independent runs. (d) Oseen-Frank elastic moduli ($K_{ii}$) predicted by DFT, with error bars estimated via a bootstrapping procedure~\cite{tortora2020chiral}. Markers denote simulation results obtained for the same molecular model using $l_p=16\,\sigma$~\cite{milchev2018nematic}. Inset: Shell-averaged alignment parameter $\alpha$ as a function of radial distance $r$. The black line correspond to the CB state ($\eta\simeq 0.26$).}
\end{figure*}

In the case of short chains ($l_c=16\,\sigma$), we observe a first spontaneous ordering transition at packing densities $\eta \simeq 0.25$, close to the I-N coexistence region predicted by DFT in the bulk (Fig.~\ref{fig2}c, green shaded area). This transition manifests through the emergence of an orientational ``wetting'' layer~\cite{vanRoij2000orientational,ivanov2013wall}, characterized by the formation of a thin nematic film at the cavity wall while the bulk of the system retains an isotropic structure (Figs.~\ref{fig1}d and~\ref{fig2}a) --- as evidenced by the growing peak of the radially-averaged alignment parameter $\alpha(r)$ near the sphere surface (Fig.~\ref{fig2}d, inset). The corresponding surface director arrangement is found to converge towards a ``cricket ball'' (CB)-like structure around $\eta\simeq 0.26$, associated with a regular tetrahedral pattern of defects with topological charge $s=1/2$, connected in pairs by the nematic field through the shortest segment of great-circle arcs (Fig.~\ref{fig2}a). Remarkably, a similar texture has been predicted theoretically by Vitelli and Nelson (VN)~\cite{vitelli2006nematic} for thin nematic shells in the limit of low splay-to-bend anisotropy, but has to our knowledge never been directly observed. DFT calculations~\cite{tortora2018incorporating} further reveal that the limited degree of surface alignment ($\alpha \gtrsim 0.7$) in the stable CB regime is associated with a weak stiffening of the bending rigidity $K_{33}$ relative to the splay modulus $K_{11}$  ($K_{11}\lesssim K_{33}$, Fig.~\ref{fig2}d), as well as a mean elastic modulus $K \equiv \sqrt{K_{11} K_{33}} \gtrsim 2\,K_0$ in slight excess of the thermal stability threshold $K_0 = 16\,k_BT/3\pi\sigma$ predicted for such tetrahedral arrangements~\cite{nelson2002toward} --- both of which fully corroborate the VN theory~\cite{vitelli2006nematic}. 
\par
Upon further increase of the concentration, we observe that these linked pairs of surface defects progressively move apart to reach antipodal positions, while the associated nematic domains simultaneously rotate to realign along a common axis. This process leads to the growth and eventual coalescence of the two domains through the fusion of unconnected defect pairs, which yields a longitudinal bipolar pattern with two $s=1$ ``hedgehog'' defects at the poles for $\eta \gtrsim 0.32$ (Fig.~\ref{fig2}b). A closer inspection of the density variations of $\alpha(r)$ (Fig.~\ref{fig2}d, inset) reveals that the destabilization of this tetrahedral surface arrangement is associated with a rapid divergence in the thickness of the wetting nematic film, which gradually takes over the entire cavity with increasing densities in the range $\eta \in [0.26,0.29]$ (Fig.~\ref{fig1}d). The quadrupolar-to-bipolar crossover observed in this regime is thus also consistent with the VN theory~\cite{vitelli2006nematic}, in which the lowest-energy state was found to progressively switch from a tetrahedral to a \mbox{(quasi-)bipolar} configuration by increasing the thickness of the nematic shell, as the two bulk disclination lines associated with the four $s=1/2$ surface defects are supplanted by more favorable 3D ``escaped'' arrangements~\cite{urbanski2017liquid}.

\subsection{Quantifying surface topological transitions} \label{sec:quantifying}

\begin{table}
\caption{Ideal zero-temperature values of the surface nematic OPs $\Lambda_i^{\rm id}$ (Eq.~\eqref{eq:q_dir}) and Glassmeier parameter $S_{\rm GM}^{\rm id}$ (Eq.~\eqref{eq:glassmeier}) for the various textures reported in this study. The $\Lambda_i^{\rm id}$ are derived in the limit of an infinitely-thin surface shell $\Omega$ ($r_s\to 0$).}
\label{table1}
\begin{center}
\begin{tabular}{|c|c|c|c|c|c|c|}
\cline{3-7}
\multicolumn{2}{c|}{} &\textbf{Figure} & $\Lambda_1^{\rm id}$ & $\Lambda_2^{\rm id}$ & $\Lambda_3^{\rm id}$ & $S_{\rm GM}^{\rm id}$ \\
\hline
\multirow{3}{*}{\textbf{Quadrupolar}} & Cricket ball & \textbf{\ref{fig2}a} & -1/3 & 1/6 & 1/6 & 1\\
\cline{2-7}
& Great circle & \textbf{\ref{fig3}a} & -1/8 & 1/16 & 1/16 & 1/2 \\
\cline{2-7}
& Tennis ball & \textbf{\ref{fig5}a} &-1/6 & -1/6 & 1/3 & 1 \\
\hline
\hline
\multirow{2}{*}{\textbf{Bipolar}} & Longitudinal & \textbf{\ref{fig2}b} & -1/8 & -1/8 & 1/4 & 0 \\
\cline{2-7}
& Latitudinal & \textbf{\ref{fig5}c} & -1/2 & 1/4 & 1/4 & 0 \\
\hline
\end{tabular}
\end{center}
\end{table}

Qualitatively, this transition from quadrupolar to bipolar surface order may be characterized by the eigenvalues $\Lambda_i$ of $\mathcal{S}(\Omega)$ (Eq.~\eqref{eq:q_dir}). For instance, elementary symmetry considerations~\cite{khadilkar2018self} lead to $\Lambda_2=\Lambda_3=-\Lambda_1/2$ for the ``cricket-ball'' texture and $\Lambda_1=\Lambda_2=-\Lambda_3/2$ for the longitudinal bipolar state (Figs.~\ref{fig2}a-b), although the actual values of the $\Lambda_i$ generally depend on both thermal fluctuations and the finite thickness $r_s$ of the surface shell $\Omega$ (c.f.~Table~\ref{table1}). In an attempt to provide a more quantitative account of such topological crossovers, several studies~\cite{nikoubashman2017semiflexible,milchev2018densely,khadilkar2018self} have introduced a tensor involving pairs of adjacent bonds,
\begin{equation}
  \label{eq:q_normal}
  \mathcal{N}_k \equiv \Big(\widehat{\mathbf{t}_k\times \mathbf{t}_{k+1}}\Big) \otimes \Big(\widehat{\mathbf{t}_k\times \mathbf{t}_{k+1}}\Big),
\end{equation}
where $\times$ and $\otimes$ respectively denote the vector cross and dyadic products. We show in the SI that in the case of semi-flexible chains, the equipartition theorem imposes
\begin{equation*}
  \big\langle \mathcal{N}_k\big\rangle = \frac{\mathcal{I}- \big\langle \mathcal{Q}_k\big\rangle}{3} + \mathcal{O}\bigg(\frac{l_b}{l_p}\bigg),
\end{equation*}
with $\mathcal{I}$ the identity tensor and $\mathcal{Q}_k$ the standard Landau-de Gennes tensor (Eq.~\eqref{eq:deGennes}), which is found to be quantitatively satisfied for all systems considered here (Figs.~S2d-f). Hence, $\big\langle\mathcal{N}_k\big\rangle$ is effectively equivalent to $\big\langle \mathcal{Q}_k\big\rangle$, and the interpretation of its eigenvalues $\mu_i$ is similarly affected by the ambiguities discussed in Sec.~\ref{appendix:describing} for the $\lambda_i$ --- suggesting that they may not reliably yield any additional insights as to the nature of these transitions. 
\par
More accurately, we here implement a simple unsupervised learning procedure to directly compute the positions of surface topological defects, which may be summarized as follows. We pick a random point $\mathbf{r}_{\rm trial}$ on the membrane surface, and consider its arbitrarily-large neighborhood $\Xi_{\rm trial} \equiv \big\{\mathbf{r}\in V \mid \lVert\mathbf{r}-\mathbf{r}_{\rm trial}\rVert < r_{\rm probe}\big\}$. We identify all polymer bonds with centers of mass $\mathbf{r}_k \in \Xi_{\rm trial}$, and compute the tensor $\mathcal{Q}_{\rm trial} \equiv \big\langle \mathcal{Q}_k\big\rangle_{\Xi_{\rm trial}}$ as in Eq.~\eqref{eq:q_loc}, from which the local degree of surface alignment may be estimated as $\alpha_{\rm trial} \equiv 2\big(\lambda_3^{\rm trial}-\lambda_2^{\rm trial}\big)/3$ (Eq.~\eqref{eq:alpha}). Since points associated with higher values of $\alpha_{\rm trial}$ are less likely to be located in the immediate vicinity of a defect, we may construct a loose collection of potential candidates by recording the position $\mathbf{r}_{\rm trial}$ with a probability $1-\alpha_{\rm trial}$, and repeating the process until a number $\mathcal{O}\big(10^4\big)$ of observations is reached. We here set $r_{\rm probe}=5\,\sigma$, although our results were found to be insensitive to any choice of $r_{\rm probe}$ such that $\sigma \ll r_{\rm probe} \ll R$.
\par
The two-fold (``head-tail'') local symmetry of the surface director field, which results from the strong tangential anchoring of the system in the nematic wetting regime (Figs.~\ref{fig1}a-c, inset), implies that the number of stable topological defects may not exceed 4 in the ground state~\cite{lubensky1992orientational}. We thus infer the most likely surface arrangements of defects in equilibrated simulations by spherical $k$-means clustering analysis~\cite{banerjee2005clustering} of the previous set of trial points, using a fixed number $k=4$ of centroids. The geometry of the defects may then be quantified through the Glassmeier parameter~\cite{robert1998tetrahedron},
\begin{equation}
  \label{eq:glassmeier}
  S_{\rm GM} = \frac{1}{2} \bigg(\frac{V_{\rm tet}}{V_{\rm reg}}+\frac{A_{\rm tet}}{A_{\rm reg}}\bigg),
\end{equation}
with $V_{\rm tet}$ and $A_{\rm tet}$ the respective volume and surface area of the tetrahedron defined by the 4 defects, rescaled by those $V_{\rm reg}$ and $A_{\rm reg}$ of the regular tetrahedron with identical circumsphere. $S_{\rm GM}=1$ characterizes regular tetrahedral order, while $S_{\rm GM}=0$ if all 4 defects are collinear --- indicating a degenerate bipolar configuration with two antipodal pairs of adjacent defects (Fig.~\ref{fig2}d). $S_{\rm GM}=1/2$ then describes the intermediary case where all defects lie coplanar in the membrane equatorial plane, which identifies the so-called ``great-circle'' configuration (c.f.~Fig.~\ref{fig3}a).
\par
To assess the reliability of the defect localization scheme, we represent in Fig.~\ref{fig5}d the density variations of $S_{\rm GM}$ against the Zhang-Chen axial OP $S_{\rm ZC}\in[0,1]$~\cite{zhang2011tennis}, which corresponds to an alternative geometrical measure of the degree of tetrahedral ``tennis ball'' order that does not rely on the determination of defect positions~\cite{zhang2011tennis}. Note that $S_{\rm ZC}$, unlike $S_{\rm GM}$, is specifically tailored to the particular director symmetries characterizing the ``tennis ball'' structure, and is generally ill-defined for the other surface assemblies reported here. Therefore, the excellent quantitative agreement between $S_{\rm GM}$ and $S_{\rm ZC}$ in Fig.~\ref{fig5}d evidences the ability of $S_{\rm GM}$ to correctly capture the evolution of the underlying director patterns in this case, and suggests its suitability as a robust and versatile metric to quantify the associated surface topological transitions.

\subsection{Splay- versus bend-dominated regimes}

\begin{figure*}[htpb]
  \includegraphics[width=2\columnwidth]{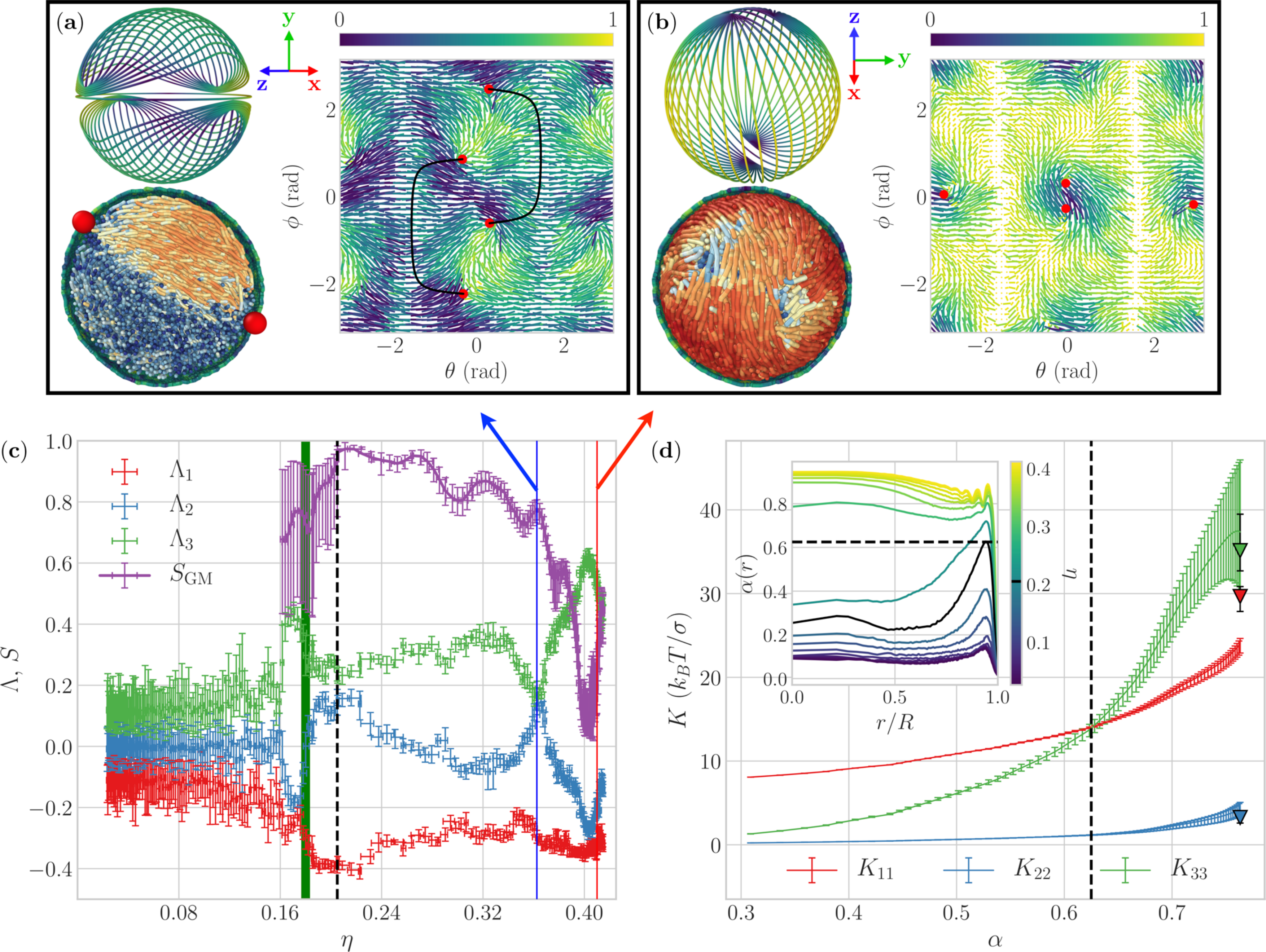}
  \caption{\label{fig3}Confined phase behavior of near-persistence-length DNA-like chains ($l_c = 32\,\sigma$). (a) The ``great-circle'' state, with $\Lambda_x=\Lambda_z=-\Lambda_y/2>0$ (Table~\ref{table1}). All defects approximately lie in the $\theta=0$ ($x$-$z$) equatorial plane. Bottom: Cut view of the simulations through the $y$-$z$ plane. Colors reflect the projection of bond orientations onto $\mathbf{z}$. Right: Stereographic map of the simulated surface director field. Black lines represent the two circular arcs connecting pairs of defects through the poles. (b) The twisted bipolar configuration, with $\Lambda_z > \Lambda_y > \Lambda_x$. The asymmetry in $\Lambda_x$ and $\Lambda_y$ reflects the spontaneous chirality of the system. Bottom: Peeled view of the simulations through the $x$-$y$ plane. (c) and (d) are as in Fig.~\ref{fig2}.}
\end{figure*}

For longer chains ($l_c=32\,\sigma$), we find that the onset of orientational wetting at $\eta \simeq 0.20$ similarly induces a CB surface texture with $S_{\rm GM}=1$ and $\Lambda_2=\Lambda_3=-\Lambda_1/2$ (Fig.~\ref{fig3}c, black line), coincident with a weak elastic anisotropy $K_{11}\lesssim K_{33}$ (Fig.~\ref{fig3}d). However, the polymers are found to display partial orientational wetting across a wider concentration range than their shorter counterparts. Indeed, increasing densities in the range $\eta\in [0.20,0.27]$ for $l_c = 32\,\sigma$ significantly enhances the degree of alignment within the wetting layer, as quantified by the height of the peaks in $\alpha(r)$ near the membrane (Fig.~\ref{fig3}d, inset), but leads to a slower increase in the layer thickness --- measured by the limited increments in the corresponding peak widths (Figs.~\ref{fig1}d-e). This effect may reflect the increasingly-weak first-order character of the I-N transition for chains with lower $l_p/l_c$~\cite{debraaf2017self}, and leads to a rapid increase of the ratio $K_{33}/K_{11}$ within the nematic wetting layer (Fig.~\ref{fig3}d), associated with a migration of the four $s=1/2$ defects towards a common equatorial plane (Fig.~\ref{fig3}a). The resulting ``great-circle'' configuration has been predicted by theory for thin nematic shells in the limit $K_{33} \gg K_{11}$~\cite{shin2008topological}, and gives rise to a bulk-ordered state featuring two distinct nematic domains for $\eta \simeq 0.36$ (Fig.~\ref{fig3}a) --- which may be recovered by cutting the bipolar configuration in Fig.~\ref{fig2}b through any plane containing the two poles, followed by a rotation of one of the hemispheres by an angle $\pi/2$~\cite{shin2008topological}. 
\par
Further increasing the density is found to lead to the gradual realignment of the two hemispheric regions. However, unlike in Fig.~\ref{fig2}b, this mutual rotation does not lead to the full coalescence of unconnected defect pairs, but rather yields four stable ``half-hedgehogs''~\cite{nelson2002toward}, associated with a finite angular mismatch between the two corresponding surface nematic domains (Fig.~\ref{fig3}b). This arrangement is strongly reminiscent of the twisted bipolar texture observed in tangentially-anchored droplets of low-molecular-weight LCs~\cite{volovik1983topological}, whose stability has been attributed to the Williams inequality~\cite{williams1986two},
\begin{equation}
  \label{eq:williams}
  K_{33} < 2.32\,\big(K_{11}-K_{22}\big),
\end{equation}
with $K_{22}$ the twist elastic modulus. We show that Eq.~\eqref{eq:williams} is violated in the high-density regime of chains with $l_c=16\,\sigma$, but holds for $l_c \geq 32\,\sigma$ (Figs.~\ref{fig4}d-f). This crossover may be attributed to the stiffening of the splay mode ($K_{11}$, Eq.~\eqref{eq:delta_K1}) for longer chains, which results in the increasing favorability of a twisted bulk structure over the large splay distortion imposed by $s=1$ hedgehog defects~\cite{williams1986two}. For $l_c=32\,\sigma$, Fig.~\ref{fig3}d yields $K_{22}/ K_{11} \simeq 0.2$ and $K_{33}/ K_{11} \simeq 1.7$ at $\eta \simeq 0.42$, from which the Williams theory predicts a twist angle at the surface of about $\SI{30}{\degree}~$\cite{williams1986two} --- in close agreement with our simulated value of $\sim\SI{25}{\degree}$ (Figs.~\ref{fig3}b and~\ref{fig6}a).

\begin{figure*}[htpb]
  \includegraphics[width=1.5\columnwidth]{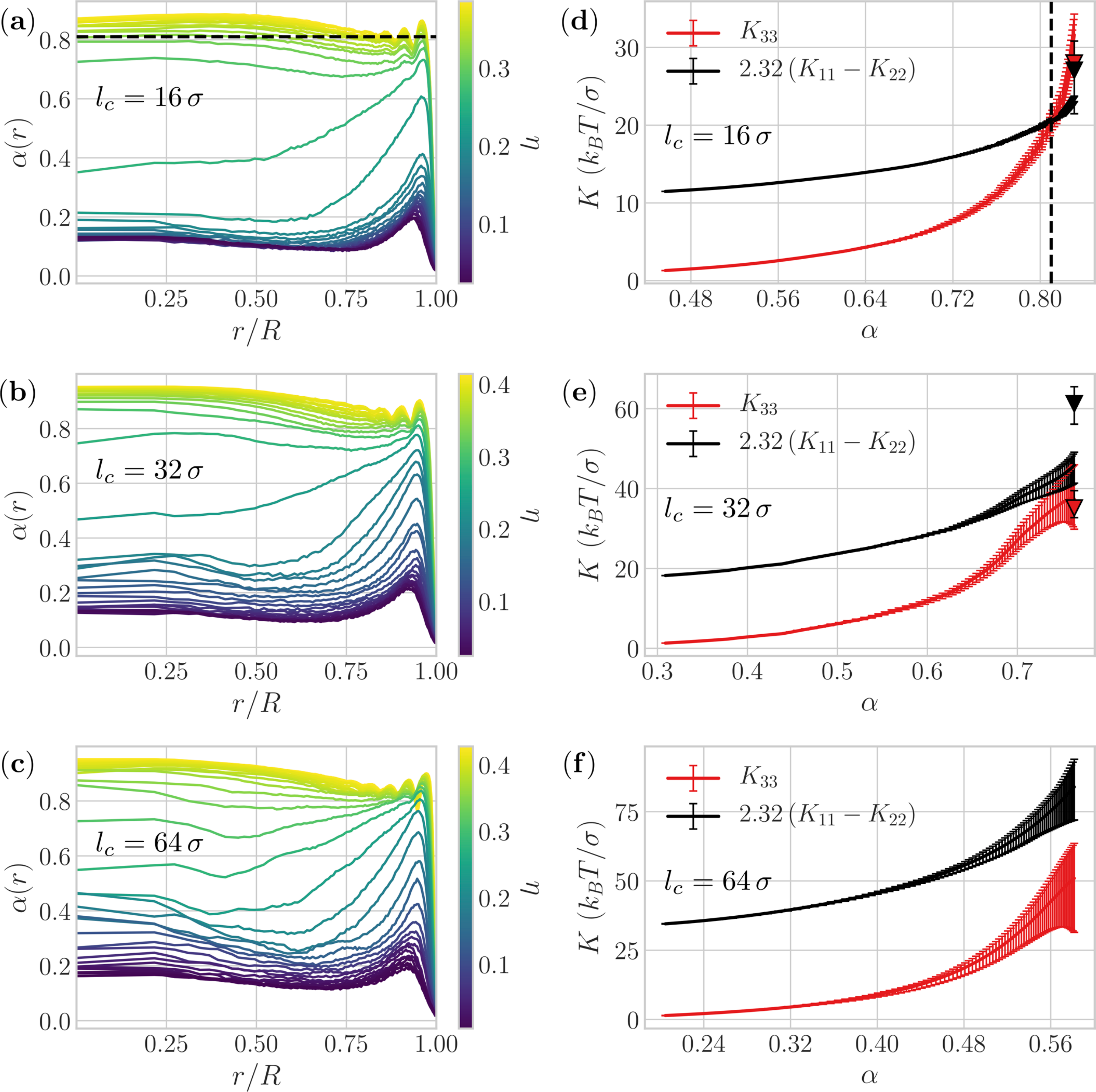}
  \caption{\label{fig4}Local alignment and elastic anisotropies of confined DNA-like chains. (a)-(c) Radial alignment parameter $\alpha(r)$ (Eq.~\eqref{eq:alpha}). (d)-(f) OF elastic anisotropies of the polymers (Eqs.~\eqref{eq:K1}--\eqref{eq:delta_K1}) as a function of the degree $\alpha$ of nematic order (Eq.~\eqref{eq:alpha_th}). The lowest reported values of $\alpha$ are obtained at the nematic binodal point~\cite{tortora2018incorporating}. Error bars are as in Fig.~\ref{fig2}d, and markers denote the simulation results of Ref.~\cite{milchev2018nematic} for $l_p=16\,\sigma$. The Williams inequality (Eq.~\eqref{eq:williams}) is satisfied at all densities for $l_c \gtrsim 32 \,\sigma$, but violated for $l_c=16\,\sigma$ in the strong alignment regime characterizing the longitudinal bipolar state ($\alpha\gtrsim 0.80$, black dashed lines in Figs.~\ref{fig4}a and d).}
\end{figure*}

Finally, for $l_c=64\,\sigma$, we find that the CB texture is replaced by a ``tennis-ball'' (TB) state in the wetting regime, obtained by linking the regular tetrahedral defect pattern pairwise through the complementary section of the same great-circle arcs (Fig.~\ref{fig5}a). We similarly impute this structure to the splay stiffening of longer chains ($K_{11}\gtrsim K_{33}$), which now promotes the bend-rich TB over the splay-rich CB arrangement (Fig.~\ref{fig5}d, inset)~\cite{vitelli2006nematic}. In this case, increasing the polymer density induces a progressive migration of the 4 defects from a regular tetrahedron ($S_{\rm GM} = 1$) towards a common meridional plane ($S_{\rm GM} \simeq 0.5$, Fig.~\ref{fig5}d). This transition is accompanied by the buckling of the two nematic domains (Fig.~\ref{fig5}b), and eventually leads to a latitudinally-ordered configuration bearing two close pairs of near-coplanar defects at the poles (Fig.~\ref{fig5}c) --- which may be viewed as a direct equivalent of half-hedgehogs~\cite{nelson2002toward} for hyperbolic $s=1$ defects.
\par
The director field is found to adopt a radially-twisted structure in the bulk, in which chains located close to the polar axis point along the north-south direction, and undergo a continuous rotation towards the latitude lines near the membrane surface with increasing axial distance (Figs.~\ref{fig5}c and~\ref{fig6}a). This arrangement, variously referred to as an escaped concentric configuration~\cite{fernandez2007topological}, twisted solenoid~\cite{shin2011filling} or condensed Hopf fibration~\cite{liang2019orientationally}, is normally only metastable in low-molecular-weight LCs~\cite{fernandez2007topological}. Here, we simply identify this texture as a special case of the twisted bipolar state (Fig.~\ref{fig3}b), in which the surface twist angle approaches \SI{90}{\degree}, and thus attribute its stability to the Williams criterion (Eq.~\eqref{eq:williams}). This inequality is expected to be quite generally satisfied for long polymers, owing to the sharp increase in the splay modulus $K_{11}$ with increasing $l_c$ (Fig.~\ref{fig5}d, inset), and would put forth the twisted latitudinal structure in Fig.~\ref{fig5}c as a robust and general template for the dense packing of long persistent chains in spherical confinement{, as discussed further below.}

\begin{figure*}[htpb]
  \includegraphics[width=2\columnwidth]{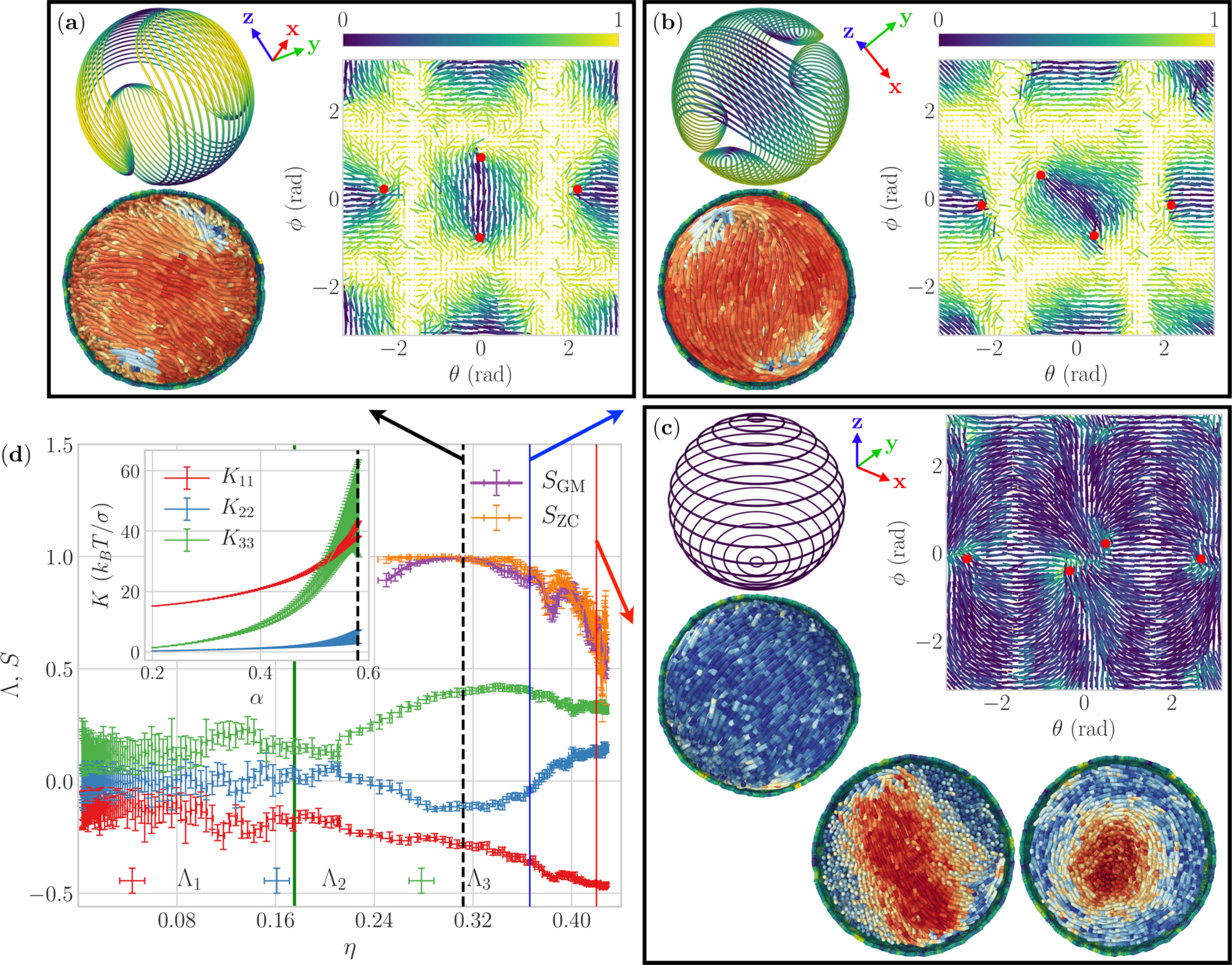}
  \caption{\label{fig5}Confined phase behavior of long DNA-like chains ($l_c = 64\,\sigma$). (a) The TB state, with $\Lambda_x=\Lambda_y=-\Lambda_z/2<0$ (Table~\ref{table1}). The TB surface texture matches the equipotential lines of the CB director field (Fig.~\ref{fig2}a)~\cite{vitelli2006nematic}. (b) The buckled TB arrangement, characterized by the migration of $s=1/2$ defects towards the $\phi=0$ ($y$-$z$) meridional plane. Note the simultaneous rotation of the domain central regions towards the $x$-$z$ equatorial lines. (c) The latitudinal bipolar configuration, with $\Lambda_x=\Lambda_y=-\Lambda_z/2>0$. Bottom: Cut view of the simulations through the $x$-$z$ (left) and $x$-$y$ (right) planes. Colors reflect the projection of bond orientations onto $\mathbf{z}$. (d) Glassmeier ($S_{\rm GM}$, Eq.~\eqref{eq:glassmeier}) and Zhang-Chen axial OP~\cite{zhang2011tennis} ($S_{\rm ZC}$) as a function of polymer packing fraction ($\eta$). Other symbols are as in Fig.~\ref{fig2}.}
\end{figure*}

{
\section{Discussion} \label{sec:discussion}

To conclude, we have extensively explored the spontaneous assembling behavior of self-avoiding, DNA-like semi-flexible filaments encapsulated within spherical biopolymeric shells. We reveal that their ordering transitions are driven by an orientational wetting phenomenon at densities close to the bulk I-N coexistence region, which is found to extend across an increasingly-wide concentration range for polymers with larger contour lengths $l_c$ (Figs.~\ref{fig1}d-f), and which we impute to the increasingly-weak first-order character of the I-N transition reported in bulk phases of SAWLCs with lower effective rigidities $l_p/l_c$~\cite{debraaf2017self}.
\par
The nematic surface layer is found to be associated with tetrahedral patterns of $s=1/2$ topological defects, which evolve towards ``escaped'' 3D arrangements bearing 2 antipodal $s=1$ defects at higher densities, as the orientational order gradually extends to the entire cavity. This observation is in agreement with the theoretical predictions of Vitelli and Nelson~\cite{vitelli2006nematic} for spherical nematic shells of increasing thickness in the one-constant approximation of nematic elasticities, which we show by means of density functional theory to be consistent with the limited degree of local alignment within the wetting layer for the various chains studied. More precisely, we report that the surface director field concomitant with the four $s=1/2$ defects adopts a splay-rich cricket-ball texture for short filaments with $l_c \lesssim 32 \,\sigma$ ($\sim 200$ base pairs), associated with a weak elastic anisotropy $K_{11} \lesssim K_{33}$ --- while longer chains with $l_c = 64 \,\sigma$ display a bend-rich tennis-ball configuration, which we attribute to a relative stiffening of the splay elasticity $K_{11} \gtrsim K_{33}$ with increasing polymer contour lengths (Fig.~\ref{fig6}b).
\par
This rigidification process chiefly arises from compressibility contributions to the splay mode (Eq.~\eqref{eq:delta_K1}), and would eventually induce the stabilization of a splay-free state in the wetting regime of very long chains, for which one expects $K_{11} \gg K_{33}$ throughout the nematic stability range~\cite{meyer1982macroscopic}. A suitable candidate would be the latitudinal bipolar surface texture depicted in Fig.~\ref{fig5}c, which features a pure bend distortion localized at the two $s=1$ hyperbolic defects~\cite{fernandez2007topological}. Such a configuration --- characterized by the presence of a disordered, isotropic core region surrounded by a spool-like arrangement of filaments near the membrane --- has indeed been recently predicted by self-consistent field theories for long SAWLCs such that $l_p \leq R$~\cite{liang2019orientationally}. This assembly was found to be stable at densities just above the bulk I-N transition point, and could therefore be similarly interpreted as an orientational wetting phenomenon --- while its absence from the phase diagram of chains with $l_p > R$~\cite{liang2019orientationally} could reflect an inhibition of wetting for stiffer filaments, due to the incompatibility between the large polymer bending rigidity and the finite curvature of the walls~\cite{groh1999fluids,trukhina2008computer}.

\begin{figure}[htpb]
  \includegraphics[width=\columnwidth]{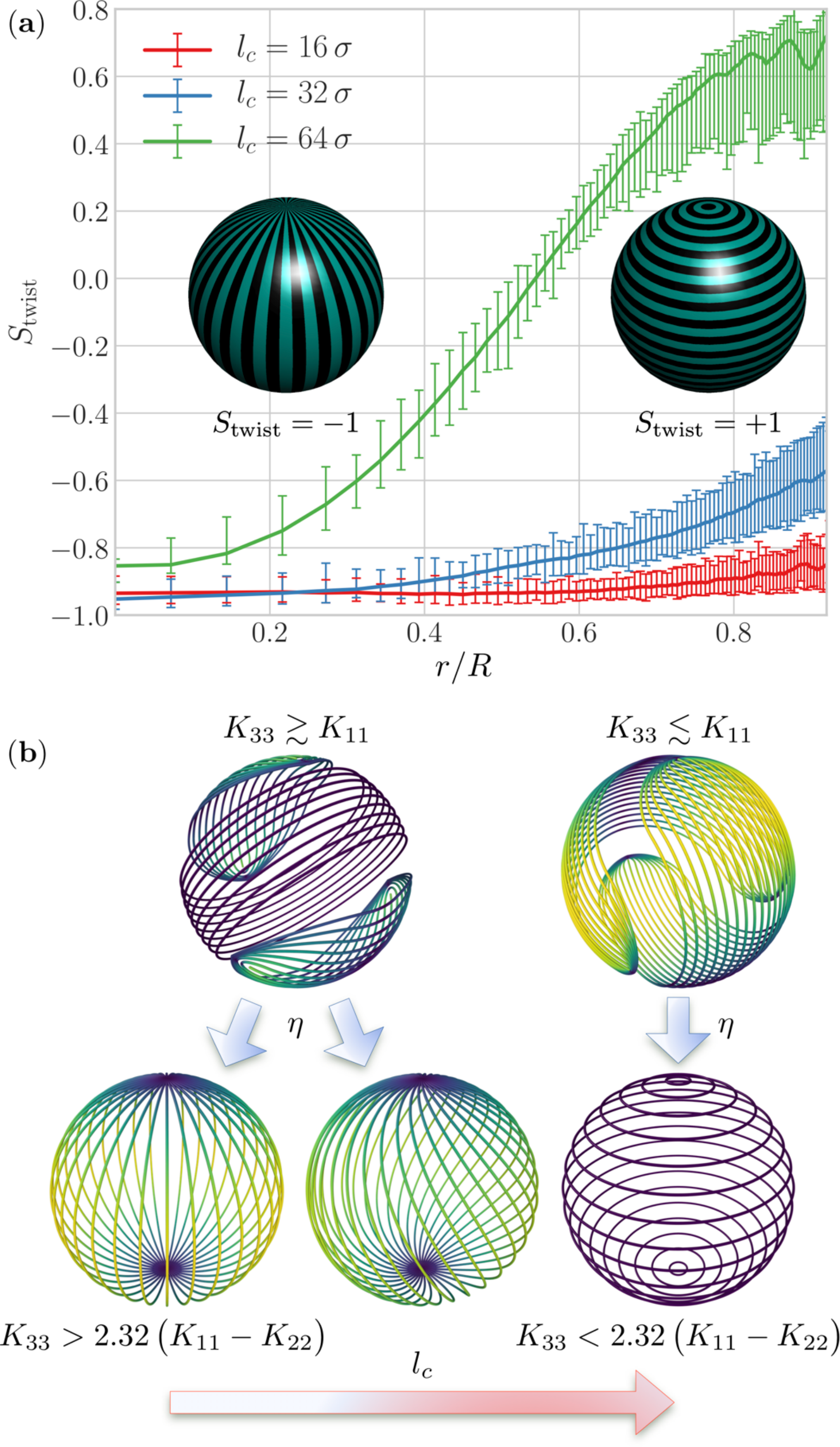}
  \caption{\label{fig6}(a) Helicoidal OP $S_{\rm twist}(r) \equiv \big\langle\cos(2\Phi_k)\big\rangle_r$~\cite{zhang2011tennis} at density $\eta \simeq 0.40$, where $\Phi_k$ is the angle between a bond vector and the local latitude circle, and we used the notations of Fig.~\ref{fig1}. $S_{\rm twist}(r)=-1$ implies that chains at radial distance $r$ point along the polar axis, and $S_{\rm twist}=1$ indicates latitudinal order. The surface value $S_{\rm twist}\simeq-0.6$ for $l_c=32\,\sigma$ specifies a finite mean twist angle $\pi/2-\Phi_k\simeq\SI{25}{\degree}$ relative to the polar axis (c.f.~Fig.~\ref{fig3}b). (b) Recapitulative phase diagram of nematic polymers in spherical confinement.}
\end{figure}

Therefore, the impact of the chain stiffness $l_p$ and finite membrane radius $R$ on the orientational wetting behavior would warrant further inquiry~\cite{milchev2021phase}. However, remarkably, our joint simulation and DFT results evidence that the phase diagram reported in Fig.~\ref{fig6}b may be fully recapitulated by continuum theories of LC elasticities independently of $R$, even though the ratio $R/l_c \in [0.4,4]$ is of order unity across all regimes investigated here (Fig.~S1). This conclusion contrasts with that of previous studies, which reported a strong dependence of macromolecular organization on confinement size due to the formation of (distorted) layered structures throughout the interior of the cavity as $R$ approaches exact multiples of $l_c$~\cite{nikoubashman2017semiflexible,milchev2018densely}. We attribute this discrepancy to our focus on more flexible, DNA-like chains ($l_p \simeq 25\,\sigma$), for which such smectic-like assemblies are expected to be significantly less favorable~\cite{salamonczyk2016smectic}. Such commensurability effects may nonetheless play a significant role in solutions of stiff, monodisperse filaments within highly-controlled geometries, as recently evidenced in strongly-confined phases of filamentous viruses~\cite{garlea2016finite}.
\par
At higher concentrations, the aforementioned stiffening of the splay mode analogously leads to a transition from a splay- (longitudinal) to a bend-dominated (latitudinal) bipolar surface texture in the bulk-ordered state as one increases $l_c$ (Fig.~\ref{fig6}b). This crossover gives rise to a growing degree of radial twist through the bulk of the phase, which may be quantitatively captured by the classical theory of Williams~\cite{williams1986two} for tangentially-anchored spherical nematic droplets (Eq.~\eqref{eq:williams}). The resulting twisted, spool-like arrangement (Fig.~\ref{fig5}c) is consistent with the folded genome structures reported in mature spherical and icosahedral bacteriophages~\cite{petrov2008packaging}, and mirrors the so-called condensed Hopf fibration/twisted solenoid conformation predicted by both continuum~\cite{shin2011filling} and self-consistent field theories~\cite{liang2019orientationally} as the general ground state of spherically-confined long chains in the high-density limit.
}
\par
Together, our findings demonstrate the ability to directly tune the topology of confined polymer solutions through the precise control of chain length and concentration, which may find significant potential applications for the rational design of patchy colloidal particles with adjustable valence and morphology~\cite{wang2016topological}. More broadly, they suggest orientational wetting as a simple interpretation for the partial ordering transitions observed during DNA ejection in bacteriophages~\cite{leforestier2010bacteriophage}, and establish the Williams criterion (Eq.~\eqref{eq:williams}) as an alternative paradigm to elucidate the chiral organization of the viral genome~\cite{marenduzzo2009dna} --- in which spontaneous twist may generically emerge from the interplay between confinement and LC elasticities, rather than specific cholesteric interactions involving the detailed molecular structure of DNA~\cite{mitov2017cholesteric}.

%------------------------------------------------

\begin{acknowledgments}
This work was supported by the Agence Nationale de la Recherche (ANR-18-CE12-0006-03, ANR-18-CE45-0022-01 \& ANR-21-CE13-0037-02). We thank the P\^ole Scientifique de Mod\'elisation Num\'erique and Centre Blaise Pascal of the ENS de Lyon for computing resources.
\par
M.M.C.T.~developed the model and theory, conducted the simulations, performed the analysis and wrote the manuscript. M.M.C.T. \& D.J. conceptualized the study and commented on the final version of the manuscript.
\end{acknowledgments}

%----------------------------------------------------------------------------------------
%	REFERENCE LIST
%----------------------------------------------------------------------------------------


%----------------------------------------------------------------------------------------


\begin{thebibliography}{82}%
\makeatletter
\providecommand \@ifxundefined [1]{%
 \@ifx{#1\undefined}
}%
\providecommand \@ifnum [1]{%
 \ifnum #1\expandafter \@firstoftwo
 \else \expandafter \@secondoftwo
 \fi
}%
\providecommand \@ifx [1]{%
 \ifx #1\expandafter \@firstoftwo
 \else \expandafter \@secondoftwo
 \fi
}%
\providecommand \natexlab [1]{#1}%
\providecommand \enquote  [1]{``#1''}%
\providecommand \bibnamefont  [1]{#1}%
\providecommand \bibfnamefont [1]{#1}%
\providecommand \citenamefont [1]{#1}%
\providecommand \href@noop [0]{\@secondoftwo}%
\providecommand \href [0]{\begingroup \@sanitize@url \@href}%
\providecommand \@href[1]{\@@startlink{#1}\@@href}%
\providecommand \@@href[1]{\endgroup#1\@@endlink}%
\providecommand \@sanitize@url [0]{\catcode `\\12\catcode `\$12\catcode
  `\&12\catcode `\#12\catcode `\^12\catcode `\_12\catcode `\%12\relax}%
\providecommand \@@startlink[1]{}%
\providecommand \@@endlink[0]{}%
\providecommand \url  [0]{\begingroup\@sanitize@url \@url }%
\providecommand \@url [1]{\endgroup\@href {#1}{\urlprefix }}%
\providecommand \urlprefix  [0]{URL }%
\providecommand \Eprint [0]{\href }%
\providecommand \doibase [0]{http://dx.doi.org/}%
\providecommand \selectlanguage [0]{\@gobble}%
\providecommand \bibinfo  [0]{\@secondoftwo}%
\providecommand \bibfield  [0]{\@secondoftwo}%
\providecommand \translation [1]{[#1]}%
\providecommand \BibitemOpen [0]{}%
\providecommand \bibitemStop [0]{}%
\providecommand \bibitemNoStop [0]{.\EOS\space}%
\providecommand \EOS [0]{\spacefactor3000\relax}%
\providecommand \BibitemShut  [1]{\csname bibitem#1\endcsname}%
\let\auto@bib@innerbib\@empty
%</preamble>
\bibitem [{\citenamefont {Ellis}(2001)}]{ellis2001macromolecular}%
  \BibitemOpen
  \bibfield  {author} {\bibinfo {author} {\bibfnamefont {R.~J.}\ \bibnamefont
  {Ellis}},\ }\href {\doibase https://doi.org/10.1016/S0968-0004(01)01938-7}
  {\bibfield  {journal} {\bibinfo  {journal} {Trends Biochem. Sci.}\ }\textbf
  {\bibinfo {volume} {26}},\ \bibinfo {pages} {597 } (\bibinfo {year}
  {2001})}\BibitemShut {NoStop}%
\bibitem [{\citenamefont {Zhou}\ \emph {et~al.}(2008)\citenamefont {Zhou},
  \citenamefont {Rivas},\ and\ \citenamefont
  {Minton}}]{zhou2008macromolecular}%
  \BibitemOpen
  \bibfield  {author} {\bibinfo {author} {\bibfnamefont {H.-X.}\ \bibnamefont
  {Zhou}}, \bibinfo {author} {\bibfnamefont {G.}~\bibnamefont {Rivas}}, \ and\
  \bibinfo {author} {\bibfnamefont {A.~P.}\ \bibnamefont {Minton}},\ }\href
  {\doibase 10.1146/annurev.biophys.37.032807.125817} {\bibfield  {journal}
  {\bibinfo  {journal} {Annu. Rev. Biophys.}\ }\textbf {\bibinfo {volume}
  {37}},\ \bibinfo {pages} {375} (\bibinfo {year} {2008})}\BibitemShut
  {NoStop}%
\bibitem [{\citenamefont {Binder}\ \emph {et~al.}(2020)\citenamefont {Binder},
  \citenamefont {Egorov}, \citenamefont {Milchev},\ and\ \citenamefont
  {Nikoubashman}}]{binder2020understanding}%
  \BibitemOpen
  \bibfield  {author} {\bibinfo {author} {\bibfnamefont {K.}~\bibnamefont
  {Binder}}, \bibinfo {author} {\bibfnamefont {S.~A.}\ \bibnamefont {Egorov}},
  \bibinfo {author} {\bibfnamefont {A.}~\bibnamefont {Milchev}}, \ and\
  \bibinfo {author} {\bibfnamefont {A.}~\bibnamefont {Nikoubashman}},\ }\href
  {\doibase 10.1088/2515-7639/ab975e} {\bibfield  {journal} {\bibinfo
  {journal} {J. Phys. Mater.}\ }\textbf {\bibinfo {volume} {3}},\ \bibinfo
  {pages} {032008} (\bibinfo {year} {2020})}\BibitemShut {NoStop}%
\bibitem [{\citenamefont {Hamley}(2010)}]{hamley2010liquid}%
  \BibitemOpen
  \bibfield  {author} {\bibinfo {author} {\bibfnamefont {I.~W.}\ \bibnamefont
  {Hamley}},\ }\href {\doibase 10.1039/B923942A} {\bibfield  {journal}
  {\bibinfo  {journal} {Soft Matter}\ }\textbf {\bibinfo {volume} {6}},\
  \bibinfo {pages} {1863} (\bibinfo {year} {2010})}\BibitemShut {NoStop}%
\bibitem [{\citenamefont {Mitov}(2017)}]{mitov2017cholesteric}%
  \BibitemOpen
  \bibfield  {author} {\bibinfo {author} {\bibfnamefont {M.}~\bibnamefont
  {Mitov}},\ }\href {\doibase 10.1039/C7SM00384F} {\bibfield  {journal}
  {\bibinfo  {journal} {Soft Matter}\ }\textbf {\bibinfo {volume} {13}},\
  \bibinfo {pages} {4176} (\bibinfo {year} {2017})}\BibitemShut {NoStop}%
\bibitem [{\citenamefont {Micheletti}\ \emph {et~al.}(2011)\citenamefont
  {Micheletti}, \citenamefont {Marenduzzo},\ and\ \citenamefont
  {Orlandini}}]{micheletti2011polymers}%
  \BibitemOpen
  \bibfield  {author} {\bibinfo {author} {\bibfnamefont {C.}~\bibnamefont
  {Micheletti}}, \bibinfo {author} {\bibfnamefont {D.}~\bibnamefont
  {Marenduzzo}}, \ and\ \bibinfo {author} {\bibfnamefont {E.}~\bibnamefont
  {Orlandini}},\ }\href {\doibase
  https://doi.org/10.1016/j.physrep.2011.03.003} {\bibfield  {journal}
  {\bibinfo  {journal} {Phys. Rep.}\ }\textbf {\bibinfo {volume} {504}},\
  \bibinfo {pages} {1 } (\bibinfo {year} {2011})}\BibitemShut {NoStop}%
\bibitem [{\citenamefont {Shaebani}\ \emph {et~al.}(2017)\citenamefont
  {Shaebani}, \citenamefont {Najafi}, \citenamefont {Farnudi}, \citenamefont
  {Bonn},\ and\ \citenamefont {Habibi}}]{shaebani2017compaction}%
  \BibitemOpen
  \bibfield  {author} {\bibinfo {author} {\bibfnamefont {M.~R.}\ \bibnamefont
  {Shaebani}}, \bibinfo {author} {\bibfnamefont {J.}~\bibnamefont {Najafi}},
  \bibinfo {author} {\bibfnamefont {A.}~\bibnamefont {Farnudi}}, \bibinfo
  {author} {\bibfnamefont {D.}~\bibnamefont {Bonn}}, \ and\ \bibinfo {author}
  {\bibfnamefont {M.}~\bibnamefont {Habibi}},\ }\href {\doibase
  10.1038/ncomms15568} {\bibfield  {journal} {\bibinfo  {journal} {Nat.
  Commun.}\ }\textbf {\bibinfo {volume} {8}},\ \bibinfo {pages} {15568}
  (\bibinfo {year} {2017})}\BibitemShut {NoStop}%
\bibitem [{\citenamefont {Curk}\ \emph {et~al.}(2019)\citenamefont {Curk},
  \citenamefont {Farrell}, \citenamefont {Dobnikar},\ and\ \citenamefont
  {Podgornik}}]{curk2019spontaneous}%
  \BibitemOpen
  \bibfield  {author} {\bibinfo {author} {\bibfnamefont {T.}~\bibnamefont
  {Curk}}, \bibinfo {author} {\bibfnamefont {J.~D.}\ \bibnamefont {Farrell}},
  \bibinfo {author} {\bibfnamefont {J.}~\bibnamefont {Dobnikar}}, \ and\
  \bibinfo {author} {\bibfnamefont {R.}~\bibnamefont {Podgornik}},\ }\href
  {\doibase 10.1103/PhysRevLett.123.047801} {\bibfield  {journal} {\bibinfo
  {journal} {Phys. Rev. Lett.}\ }\textbf {\bibinfo {volume} {123}},\ \bibinfo
  {pages} {047801} (\bibinfo {year} {2019})}\BibitemShut {NoStop}%
\bibitem [{\citenamefont {Frankel}(2003)}]{frankel2003geometry}%
  \BibitemOpen
  \bibfield  {author} {\bibinfo {author} {\bibfnamefont {T.}~\bibnamefont
  {Frankel}},\ }\href@noop {} {\emph {\bibinfo {title} {The Geometry of
  Physics: An Introduction}}},\ \bibinfo {edition} {2nd}\ ed.\ (\bibinfo
  {publisher} {Cambridge University Press},\ \bibinfo {address} {Cambridge},\
  \bibinfo {year} {2003})\BibitemShut {NoStop}%
\bibitem [{\citenamefont {Kleman}\ and\ \citenamefont
  {Lavrentovich}(2003)}]{kleman2003topological}%
  \BibitemOpen
  \bibinfo {editor} {\bibfnamefont {M.}~\bibnamefont {Kleman}}\ and\ \bibinfo
  {editor} {\bibfnamefont {O.~D.}\ \bibnamefont {Lavrentovich}},\ eds.,\
  \enquote {\bibinfo {title} {Topological theory of defects},}\ in\ \href
  {\doibase 10.1007/978-0-387-21759-8_12} {\emph {\bibinfo {booktitle} {Soft
  Matter Physics: An Introduction}}}\ (\bibinfo  {publisher} {Springer New
  York},\ \bibinfo {address} {New York, NY},\ \bibinfo {year} {2003})\ pp.\
  \bibinfo {pages} {434--471}\BibitemShut {NoStop}%
\bibitem [{\citenamefont {Urbanski}\ \emph {et~al.}(2017)\citenamefont
  {Urbanski}, \citenamefont {Reyes}, \citenamefont {Noh}, \citenamefont
  {Sharma}, \citenamefont {Geng}, \citenamefont {Jampani},\ and\ \citenamefont
  {Lagerwall}}]{urbanski2017liquid}%
  \BibitemOpen
  \bibfield  {author} {\bibinfo {author} {\bibfnamefont {M.}~\bibnamefont
  {Urbanski}}, \bibinfo {author} {\bibfnamefont {C.~G.}\ \bibnamefont {Reyes}},
  \bibinfo {author} {\bibfnamefont {J.}~\bibnamefont {Noh}}, \bibinfo {author}
  {\bibfnamefont {A.}~\bibnamefont {Sharma}}, \bibinfo {author} {\bibfnamefont
  {Y.}~\bibnamefont {Geng}}, \bibinfo {author} {\bibfnamefont {V.~S.~R.}\
  \bibnamefont {Jampani}}, \ and\ \bibinfo {author} {\bibfnamefont {J.~P.~F.}\
  \bibnamefont {Lagerwall}},\ }\href {\doibase 10.1088/1361-648x/aa5706}
  {\bibfield  {journal} {\bibinfo  {journal} {J. Phys. Condens. Matter}\
  }\textbf {\bibinfo {volume} {29}},\ \bibinfo {pages} {133003} (\bibinfo
  {year} {2017})}\BibitemShut {NoStop}%
\bibitem [{\citenamefont {Volovik}\ and\ \citenamefont
  {Lavrentovich}(1983)}]{volovik1983topological}%
  \BibitemOpen
  \bibfield  {author} {\bibinfo {author} {\bibfnamefont {G.~E.}\ \bibnamefont
  {Volovik}}\ and\ \bibinfo {author} {\bibfnamefont {O.~D.}\ \bibnamefont
  {Lavrentovich}},\ }\href {http://jetp.ac.ru/cgi-bin/dn/e_058_06_1159.pdf}
  {\bibfield  {journal} {\bibinfo  {journal} {Zh. Eksp. Teor. Fiz.}\ }\textbf
  {\bibinfo {volume} {85}},\ \bibinfo {pages} {1997} (\bibinfo {year}
  {1983})}\BibitemShut {NoStop}%
\bibitem [{\citenamefont {Lavrentovich}(1998)}]{lavrentovich1998topological}%
  \BibitemOpen
  \bibfield  {author} {\bibinfo {author} {\bibfnamefont {O.~D.}\ \bibnamefont
  {Lavrentovich}},\ }\href {\doibase 10.1080/026782998207640} {\bibfield
  {journal} {\bibinfo  {journal} {Liq. Cryst.}\ }\textbf {\bibinfo {volume}
  {24}},\ \bibinfo {pages} {117} (\bibinfo {year} {1998})}\BibitemShut
  {NoStop}%
\bibitem [{\citenamefont {Kleman}\ and\ \citenamefont
  {Lavrentovich}(2006)}]{kleman2006topological}%
  \BibitemOpen
  \bibfield  {author} {\bibinfo {author} {\bibfnamefont {M.}~\bibnamefont
  {Kleman}}\ and\ \bibinfo {author} {\bibfnamefont {O.~D.}\ \bibnamefont
  {Lavrentovich}},\ }\href {\doibase 10.1080/14786430600593016} {\bibfield
  {journal} {\bibinfo  {journal} {Phil. Mag.}\ }\textbf {\bibinfo {volume}
  {86}},\ \bibinfo {pages} {4117} (\bibinfo {year} {2006})}\BibitemShut
  {NoStop}%
\bibitem [{\citenamefont {Lopez-Leon}\ \emph {et~al.}(2011)\citenamefont
  {Lopez-Leon}, \citenamefont {Koning}, \citenamefont {Devaiah}, \citenamefont
  {Vitelli},\ and\ \citenamefont {Fernandez-Nieves}}]{lopez2011frustrated}%
  \BibitemOpen
  \bibfield  {author} {\bibinfo {author} {\bibfnamefont {T.}~\bibnamefont
  {Lopez-Leon}}, \bibinfo {author} {\bibfnamefont {V.}~\bibnamefont {Koning}},
  \bibinfo {author} {\bibfnamefont {K.~B.~S.}\ \bibnamefont {Devaiah}},
  \bibinfo {author} {\bibfnamefont {V.}~\bibnamefont {Vitelli}}, \ and\
  \bibinfo {author} {\bibfnamefont {A.}~\bibnamefont {Fernandez-Nieves}},\
  }\href {\doibase 10.1038/nphys1920} {\bibfield  {journal} {\bibinfo
  {journal} {Nat. Phys.}\ }\textbf {\bibinfo {volume} {7}},\ \bibinfo {pages}
  {391} (\bibinfo {year} {2011})}\BibitemShut {NoStop}%
\bibitem [{\citenamefont {Tomar}\ \emph {et~al.}(2012)\citenamefont {Tomar},
  \citenamefont {Hernández}, \citenamefont {Abbott}, \citenamefont
  {Hernández-Ortiz},\ and\ \citenamefont {de~Pablo}}]{tomar2012morphological}%
  \BibitemOpen
  \bibfield  {author} {\bibinfo {author} {\bibfnamefont {V.}~\bibnamefont
  {Tomar}}, \bibinfo {author} {\bibfnamefont {S.~I.}\ \bibnamefont
  {Hernández}}, \bibinfo {author} {\bibfnamefont {N.~L.}\ \bibnamefont
  {Abbott}}, \bibinfo {author} {\bibfnamefont {J.~P.}\ \bibnamefont
  {Hernández-Ortiz}}, \ and\ \bibinfo {author} {\bibfnamefont {J.~J.}\
  \bibnamefont {de~Pablo}},\ }\href {\doibase 10.1039/C2SM25383F} {\bibfield
  {journal} {\bibinfo  {journal} {Soft Matter}\ }\textbf {\bibinfo {volume}
  {8}},\ \bibinfo {pages} {8679} (\bibinfo {year} {2012})}\BibitemShut
  {NoStop}%
\bibitem [{\citenamefont {Chiccoli}\ \emph {et~al.}(1990)\citenamefont
  {Chiccoli}, \citenamefont {Pasini}, \citenamefont {Semeria},\ and\
  \citenamefont {Zannoni}}]{chiccoli1990computer}%
  \BibitemOpen
  \bibfield  {author} {\bibinfo {author} {\bibfnamefont {C.}~\bibnamefont
  {Chiccoli}}, \bibinfo {author} {\bibfnamefont {P.}~\bibnamefont {Pasini}},
  \bibinfo {author} {\bibfnamefont {F.}~\bibnamefont {Semeria}}, \ and\
  \bibinfo {author} {\bibfnamefont {C.}~\bibnamefont {Zannoni}},\ }\href
  {\doibase https://doi.org/10.1016/0375-9601(90)90103-U} {\bibfield  {journal}
  {\bibinfo  {journal} {Phys. Lett. A}\ }\textbf {\bibinfo {volume} {150}},\
  \bibinfo {pages} {311} (\bibinfo {year} {1990})}\BibitemShut {NoStop}%
\bibitem [{\citenamefont {Chiccoli}\ \emph {et~al.}(1997)\citenamefont
  {Chiccoli}, \citenamefont {Lavrentovich}, \citenamefont {Pasini},\ and\
  \citenamefont {Zannoni}}]{chiccoli1997monte}%
  \BibitemOpen
  \bibfield  {author} {\bibinfo {author} {\bibfnamefont {C.}~\bibnamefont
  {Chiccoli}}, \bibinfo {author} {\bibfnamefont {O.~D.}\ \bibnamefont
  {Lavrentovich}}, \bibinfo {author} {\bibfnamefont {P.}~\bibnamefont
  {Pasini}}, \ and\ \bibinfo {author} {\bibfnamefont {C.}~\bibnamefont
  {Zannoni}},\ }\href {\doibase 10.1103/PhysRevLett.79.4401} {\bibfield
  {journal} {\bibinfo  {journal} {Phys. Rev. Lett.}\ }\textbf {\bibinfo
  {volume} {79}},\ \bibinfo {pages} {4401} (\bibinfo {year}
  {1997})}\BibitemShut {NoStop}%
\bibitem [{\citenamefont {Brada\v{c}}\ \emph {et~al.}(1998)\citenamefont
  {Brada\v{c}}, \citenamefont {Kralj},\ and\ \citenamefont
  {\v{Z}umer}}]{bradac1998molecular}%
  \BibitemOpen
  \bibfield  {author} {\bibinfo {author} {\bibfnamefont {Z.}~\bibnamefont
  {Brada\v{c}}}, \bibinfo {author} {\bibfnamefont {S.}~\bibnamefont {Kralj}}, \
  and\ \bibinfo {author} {\bibfnamefont {S.}~\bibnamefont {\v{Z}umer}},\ }\href
  {\doibase 10.1103/PhysRevE.58.7447} {\bibfield  {journal} {\bibinfo
  {journal} {Phys. Rev. E}\ }\textbf {\bibinfo {volume} {58}},\ \bibinfo
  {pages} {7447} (\bibinfo {year} {1998})}\BibitemShut {NoStop}%
\bibitem [{\citenamefont {Stark}\ \emph {et~al.}(1999)\citenamefont {Stark},
  \citenamefont {Stelzer},\ and\ \citenamefont {Bernhard}}]{stark1999water}%
  \BibitemOpen
  \bibfield  {author} {\bibinfo {author} {\bibfnamefont {H.}~\bibnamefont
  {Stark}}, \bibinfo {author} {\bibfnamefont {J.}~\bibnamefont {Stelzer}}, \
  and\ \bibinfo {author} {\bibfnamefont {R.}~\bibnamefont {Bernhard}},\ }\href
  {\doibase 10.1007/s100510050881} {\bibfield  {journal} {\bibinfo  {journal}
  {Eur. Phys. J. B}\ }\textbf {\bibinfo {volume} {10}},\ \bibinfo {pages} {515}
  (\bibinfo {year} {1999})}\BibitemShut {NoStop}%
\bibitem [{\citenamefont {Andrienko}\ \emph {et~al.}(2002)\citenamefont
  {Andrienko}, \citenamefont {Allen}, \citenamefont {Ska\v{c}ej},\ and\
  \citenamefont {\v{Z}umer}}]{andrienko2002defect}%
  \BibitemOpen
  \bibfield  {author} {\bibinfo {author} {\bibfnamefont {D.}~\bibnamefont
  {Andrienko}}, \bibinfo {author} {\bibfnamefont {M.~P.}\ \bibnamefont
  {Allen}}, \bibinfo {author} {\bibfnamefont {G.}~\bibnamefont {Ska\v{c}ej}}, \
  and\ \bibinfo {author} {\bibfnamefont {S.}~\bibnamefont {\v{Z}umer}},\ }\href
  {\doibase 10.1103/PhysRevE.65.041702} {\bibfield  {journal} {\bibinfo
  {journal} {Phys. Rev. E}\ }\textbf {\bibinfo {volume} {65}},\ \bibinfo
  {pages} {041702} (\bibinfo {year} {2002})}\BibitemShut {NoStop}%
\bibitem [{\citenamefont {Guzm\'an}\ \emph {et~al.}(2003)\citenamefont
  {Guzm\'an}, \citenamefont {Kim}, \citenamefont {Grollau}, \citenamefont
  {Abbott},\ and\ \citenamefont {de~Pablo}}]{guzman2003defect}%
  \BibitemOpen
  \bibfield  {author} {\bibinfo {author} {\bibfnamefont {O.}~\bibnamefont
  {Guzm\'an}}, \bibinfo {author} {\bibfnamefont {E.~B.}\ \bibnamefont {Kim}},
  \bibinfo {author} {\bibfnamefont {S.}~\bibnamefont {Grollau}}, \bibinfo
  {author} {\bibfnamefont {N.~L.}\ \bibnamefont {Abbott}}, \ and\ \bibinfo
  {author} {\bibfnamefont {J.~J.}\ \bibnamefont {de~Pablo}},\ }\href {\doibase
  10.1103/PhysRevLett.91.235507} {\bibfield  {journal} {\bibinfo  {journal}
  {Phys. Rev. Lett.}\ }\textbf {\bibinfo {volume} {91}},\ \bibinfo {pages}
  {235507} (\bibinfo {year} {2003})}\BibitemShut {NoStop}%
\bibitem [{\citenamefont {Gharbi}\ \emph {et~al.}(2013)\citenamefont {Gharbi},
  \citenamefont {Seč}, \citenamefont {Lopez-Leon}, \citenamefont {Nobili},
  \citenamefont {Ravnik}, \citenamefont {Žumer},\ and\ \citenamefont
  {Blanc}}]{gharbi2013microparticles}%
  \BibitemOpen
  \bibfield  {author} {\bibinfo {author} {\bibfnamefont {M.~A.}\ \bibnamefont
  {Gharbi}}, \bibinfo {author} {\bibfnamefont {D.}~\bibnamefont {Seč}},
  \bibinfo {author} {\bibfnamefont {T.}~\bibnamefont {Lopez-Leon}}, \bibinfo
  {author} {\bibfnamefont {M.}~\bibnamefont {Nobili}}, \bibinfo {author}
  {\bibfnamefont {M.}~\bibnamefont {Ravnik}}, \bibinfo {author} {\bibfnamefont
  {S.}~\bibnamefont {Žumer}}, \ and\ \bibinfo {author} {\bibfnamefont
  {C.}~\bibnamefont {Blanc}},\ }\href {\doibase 10.1039/C3SM00126A} {\bibfield
  {journal} {\bibinfo  {journal} {Soft Matter}\ }\textbf {\bibinfo {volume}
  {9}},\ \bibinfo {pages} {6911} (\bibinfo {year} {2013})}\BibitemShut
  {NoStop}%
\bibitem [{\citenamefont {de~Gennes}\ and\ \citenamefont
  {Prost}(1993)}]{deGennes1993physics}%
  \BibitemOpen
  \bibfield  {author} {\bibinfo {author} {\bibfnamefont {P.-G.}\ \bibnamefont
  {de~Gennes}}\ and\ \bibinfo {author} {\bibfnamefont {J.}~\bibnamefont
  {Prost}},\ }\href@noop {} {\emph {\bibinfo {title} {The Physics of Liquid
  Crystals}}},\ \bibinfo {edition} {2nd}\ ed.\ (\bibinfo  {publisher}
  {Clarendon Press},\ \bibinfo {address} {Oxford},\ \bibinfo {year}
  {1993})\BibitemShut {NoStop}%
\bibitem [{\citenamefont {Nikoubashman}\ \emph {et~al.}(2017)\citenamefont
  {Nikoubashman}, \citenamefont {Vega}, \citenamefont {Binder},\ and\
  \citenamefont {Milchev}}]{nikoubashman2017semiflexible}%
  \BibitemOpen
  \bibfield  {author} {\bibinfo {author} {\bibfnamefont {A.}~\bibnamefont
  {Nikoubashman}}, \bibinfo {author} {\bibfnamefont {D.~A.}\ \bibnamefont
  {Vega}}, \bibinfo {author} {\bibfnamefont {K.}~\bibnamefont {Binder}}, \ and\
  \bibinfo {author} {\bibfnamefont {A.}~\bibnamefont {Milchev}},\ }\href
  {\doibase 10.1103/PhysRevLett.118.217803} {\bibfield  {journal} {\bibinfo
  {journal} {Phys. Rev. Lett.}\ }\textbf {\bibinfo {volume} {118}},\ \bibinfo
  {pages} {217803} (\bibinfo {year} {2017})}\BibitemShut {NoStop}%
\bibitem [{\citenamefont {Milchev}\ \emph
  {et~al.}(2018{\natexlab{a}})\citenamefont {Milchev}, \citenamefont {Egorov},
  \citenamefont {Vega}, \citenamefont {Binder},\ and\ \citenamefont
  {Nikoubashman}}]{milchev2018densely}%
  \BibitemOpen
  \bibfield  {author} {\bibinfo {author} {\bibfnamefont {A.}~\bibnamefont
  {Milchev}}, \bibinfo {author} {\bibfnamefont {S.~A.}\ \bibnamefont {Egorov}},
  \bibinfo {author} {\bibfnamefont {D.~A.}\ \bibnamefont {Vega}}, \bibinfo
  {author} {\bibfnamefont {K.}~\bibnamefont {Binder}}, \ and\ \bibinfo {author}
  {\bibfnamefont {A.}~\bibnamefont {Nikoubashman}},\ }\href {\doibase
  10.1021/acs.macromol.7b02643} {\bibfield  {journal} {\bibinfo  {journal}
  {Macromolecules}\ }\textbf {\bibinfo {volume} {51}},\ \bibinfo {pages} {2002}
  (\bibinfo {year} {2018}{\natexlab{a}})}\BibitemShut {NoStop}%
\bibitem [{\citenamefont {Khadilkar}\ and\ \citenamefont
  {Nikoubashman}(2018)}]{khadilkar2018self}%
  \BibitemOpen
  \bibfield  {author} {\bibinfo {author} {\bibfnamefont {M.~R.}\ \bibnamefont
  {Khadilkar}}\ and\ \bibinfo {author} {\bibfnamefont {A.}~\bibnamefont
  {Nikoubashman}},\ }\href {\doibase 10.1039/C8SM01170B} {\bibfield  {journal}
  {\bibinfo  {journal} {Soft Matter}\ }\textbf {\bibinfo {volume} {14}},\
  \bibinfo {pages} {6903} (\bibinfo {year} {2018})}\BibitemShut {NoStop}%
\bibitem [{\citenamefont {Nikoubashman}(2021)}]{nikoubashman2021ordering}%
  \BibitemOpen
  \bibfield  {author} {\bibinfo {author} {\bibfnamefont {A.}~\bibnamefont
  {Nikoubashman}},\ }\href {\doibase 10.1063/5.0038052} {\bibfield  {journal}
  {\bibinfo  {journal} {J. Chem. Phys.}\ }\textbf {\bibinfo {volume} {154}},\
  \bibinfo {pages} {090901} (\bibinfo {year} {2021})},\ \Eprint
  {http://arxiv.org/abs/https://doi.org/10.1063/5.0038052}
  {https://doi.org/10.1063/5.0038052} \BibitemShut {NoStop}%
\bibitem [{\citenamefont {Marenduzzo}\ \emph {et~al.}(2010)\citenamefont
  {Marenduzzo}, \citenamefont {Micheletti},\ and\ \citenamefont
  {Orlandini}}]{marenduzzo2010biopolymer}%
  \BibitemOpen
  \bibfield  {author} {\bibinfo {author} {\bibfnamefont {D.}~\bibnamefont
  {Marenduzzo}}, \bibinfo {author} {\bibfnamefont {C.}~\bibnamefont
  {Micheletti}}, \ and\ \bibinfo {author} {\bibfnamefont {E.}~\bibnamefont
  {Orlandini}},\ }\href {\doibase 10.1088/0953-8984/22/28/283102} {\bibfield
  {journal} {\bibinfo  {journal} {J. Phys. Condens. Matter}\ }\textbf {\bibinfo
  {volume} {22}},\ \bibinfo {pages} {283102} (\bibinfo {year}
  {2010})}\BibitemShut {NoStop}%
\bibitem [{\citenamefont {Shin}\ and\ \citenamefont
  {Grason}(2011)}]{shin2011filling}%
  \BibitemOpen
  \bibfield  {author} {\bibinfo {author} {\bibfnamefont {H.}~\bibnamefont
  {Shin}}\ and\ \bibinfo {author} {\bibfnamefont {G.~M.}\ \bibnamefont
  {Grason}},\ }\href {\doibase 10.1209/0295-5075/96/36007} {\bibfield
  {journal} {\bibinfo  {journal} {EPL}\ }\textbf {\bibinfo {volume} {96}},\
  \bibinfo {pages} {36007} (\bibinfo {year} {2011})}\BibitemShut {NoStop}%
\bibitem [{\citenamefont {Zhang}\ and\ \citenamefont
  {Chen}(2011)}]{zhang2011tennis}%
  \BibitemOpen
  \bibfield  {author} {\bibinfo {author} {\bibfnamefont {W.-Y.}\ \bibnamefont
  {Zhang}}\ and\ \bibinfo {author} {\bibfnamefont {J.~Z.~Y.}\ \bibnamefont
  {Chen}},\ }\href {\doibase 10.1209/0295-5075/94/43001} {\bibfield  {journal}
  {\bibinfo  {journal} {EPL}\ }\textbf {\bibinfo {volume} {94}},\ \bibinfo
  {pages} {43001} (\bibinfo {year} {2011})}\BibitemShut {NoStop}%
\bibitem [{\citenamefont {Chen}(2016)}]{chen2016theory}%
  \BibitemOpen
  \bibfield  {author} {\bibinfo {author} {\bibfnamefont {J.~Z.~Y.}\
  \bibnamefont {Chen}},\ }\href {\doibase
  https://doi.org/10.1016/j.progpolymsci.2015.09.002} {\bibfield  {journal}
  {\bibinfo  {journal} {Prog. Polym. Sci.}\ }\textbf {\bibinfo {volume}
  {54-55}},\ \bibinfo {pages} {3 } (\bibinfo {year} {2016})}\BibitemShut
  {NoStop}%
\bibitem [{\citenamefont {Liang}\ \emph {et~al.}(2019)\citenamefont {Liang},
  \citenamefont {Jiang},\ and\ \citenamefont
  {Chen}}]{liang2019orientationally}%
  \BibitemOpen
  \bibfield  {author} {\bibinfo {author} {\bibfnamefont {Q.}~\bibnamefont
  {Liang}}, \bibinfo {author} {\bibfnamefont {Y.}~\bibnamefont {Jiang}}, \ and\
  \bibinfo {author} {\bibfnamefont {J.~Z.~Y.}\ \bibnamefont {Chen}},\ }\href
  {\doibase 10.1103/PhysRevE.100.032502} {\bibfield  {journal} {\bibinfo
  {journal} {Phys. Rev. E}\ }\textbf {\bibinfo {volume} {100}},\ \bibinfo
  {pages} {032502} (\bibinfo {year} {2019})}\BibitemShut {NoStop}%
\bibitem [{\citenamefont {G{\^a}rlea}\ \emph {et~al.}(2016)\citenamefont
  {G{\^a}rlea}, \citenamefont {Mulder}, \citenamefont {Alvarado}, \citenamefont
  {Dammone}, \citenamefont {Aarts}, \citenamefont {Lettinga}, \citenamefont
  {Koenderink},\ and\ \citenamefont {Mulder}}]{garlea2016finite}%
  \BibitemOpen
  \bibfield  {author} {\bibinfo {author} {\bibfnamefont {I.~C.}\ \bibnamefont
  {G{\^a}rlea}}, \bibinfo {author} {\bibfnamefont {P.}~\bibnamefont {Mulder}},
  \bibinfo {author} {\bibfnamefont {J.}~\bibnamefont {Alvarado}}, \bibinfo
  {author} {\bibfnamefont {O.}~\bibnamefont {Dammone}}, \bibinfo {author}
  {\bibfnamefont {D.~G. A.~L.}\ \bibnamefont {Aarts}}, \bibinfo {author}
  {\bibfnamefont {M.~P.}\ \bibnamefont {Lettinga}}, \bibinfo {author}
  {\bibfnamefont {G.~H.}\ \bibnamefont {Koenderink}}, \ and\ \bibinfo {author}
  {\bibfnamefont {B.~M.}\ \bibnamefont {Mulder}},\ }\href {\doibase
  10.1038/ncomms12112} {\bibfield  {journal} {\bibinfo  {journal} {Nat.
  Commun.}\ }\textbf {\bibinfo {volume} {7}},\ \bibinfo {pages} {12112}
  (\bibinfo {year} {2016})}\BibitemShut {NoStop}%
\bibitem [{\citenamefont {Broedersz}\ and\ \citenamefont
  {MacKintosh}(2014)}]{broedersz2014modeling}%
  \BibitemOpen
  \bibfield  {author} {\bibinfo {author} {\bibfnamefont {C.~P.}\ \bibnamefont
  {Broedersz}}\ and\ \bibinfo {author} {\bibfnamefont {F.~C.}\ \bibnamefont
  {MacKintosh}},\ }\href {\doibase 10.1103/RevModPhys.86.995} {\bibfield
  {journal} {\bibinfo  {journal} {Rev. Mod. Phys.}\ }\textbf {\bibinfo {volume}
  {86}},\ \bibinfo {pages} {995} (\bibinfo {year} {2014})}\BibitemShut
  {NoStop}%
\bibitem [{\citenamefont {Onsager}(1949)}]{onsager1949effects}%
  \BibitemOpen
  \bibfield  {author} {\bibinfo {author} {\bibfnamefont {L.}~\bibnamefont
  {Onsager}},\ }\href {\doibase 10.1111/j.1749-6632.1949.tb27296.x} {\bibfield
  {journal} {\bibinfo  {journal} {Ann. N.Y. Acad. Sci.}\ }\textbf {\bibinfo
  {volume} {51}},\ \bibinfo {pages} {627} (\bibinfo {year} {1949})}\BibitemShut
  {NoStop}%
\bibitem [{\citenamefont {Grest}\ and\ \citenamefont
  {Kremer}(1986)}]{grest1986molecular}%
  \BibitemOpen
  \bibfield  {author} {\bibinfo {author} {\bibfnamefont {G.~S.}\ \bibnamefont
  {Grest}}\ and\ \bibinfo {author} {\bibfnamefont {K.}~\bibnamefont {Kremer}},\
  }\href {\doibase 10.1103/PhysRevA.33.3628} {\bibfield  {journal} {\bibinfo
  {journal} {Phys. Rev. A}\ }\textbf {\bibinfo {volume} {33}},\ \bibinfo
  {pages} {3628} (\bibinfo {year} {1986})}\BibitemShut {NoStop}%
\bibitem [{\citenamefont {Kremer}\ and\ \citenamefont
  {Grest}(1990)}]{kremer1990dynamics}%
  \BibitemOpen
  \bibfield  {author} {\bibinfo {author} {\bibfnamefont {K.}~\bibnamefont
  {Kremer}}\ and\ \bibinfo {author} {\bibfnamefont {G.~S.}\ \bibnamefont
  {Grest}},\ }\href {\doibase 10.1063/1.458541} {\bibfield  {journal} {\bibinfo
   {journal} {J. Chem. Phys.}\ }\textbf {\bibinfo {volume} {92}},\ \bibinfo
  {pages} {5057} (\bibinfo {year} {1990})}\BibitemShut {NoStop}%
\bibitem [{\citenamefont {Kratky}\ and\ \citenamefont
  {Porod}(1949)}]{kratky1949diffuse}%
  \BibitemOpen
  \bibfield  {author} {\bibinfo {author} {\bibfnamefont {O.}~\bibnamefont
  {Kratky}}\ and\ \bibinfo {author} {\bibfnamefont {G.}~\bibnamefont {Porod}},\
  }\href {\doibase https://doi.org/10.1016/0095-8522(49)90032-X} {\bibfield
  {journal} {\bibinfo  {journal} {J. Colloid Sci.}\ }\textbf {\bibinfo {volume}
  {4}},\ \bibinfo {pages} {35} (\bibinfo {year} {1949})}\BibitemShut {NoStop}%
\bibitem [{\citenamefont {Weeks}\ \emph {et~al.}(1971)\citenamefont {Weeks},
  \citenamefont {Chandler},\ and\ \citenamefont {Andersen}}]{weeks1971role}%
  \BibitemOpen
  \bibfield  {author} {\bibinfo {author} {\bibfnamefont {J.~D.}\ \bibnamefont
  {Weeks}}, \bibinfo {author} {\bibfnamefont {D.}~\bibnamefont {Chandler}}, \
  and\ \bibinfo {author} {\bibfnamefont {H.~C.}\ \bibnamefont {Andersen}},\
  }\href {\doibase 10.1063/1.1674820} {\bibfield  {journal} {\bibinfo
  {journal} {J. Chem. Phys.}\ }\textbf {\bibinfo {volume} {54}},\ \bibinfo
  {pages} {5237} (\bibinfo {year} {1971})}\BibitemShut {NoStop}%
\bibitem [{\citenamefont {Egorov}\ \emph {et~al.}(2016)\citenamefont {Egorov},
  \citenamefont {Milchev}, \citenamefont {Virnau},\ and\ \citenamefont
  {Binder}}]{egorov2016insight}%
  \BibitemOpen
  \bibfield  {author} {\bibinfo {author} {\bibfnamefont {S.~A.}\ \bibnamefont
  {Egorov}}, \bibinfo {author} {\bibfnamefont {A.}~\bibnamefont {Milchev}},
  \bibinfo {author} {\bibfnamefont {P.}~\bibnamefont {Virnau}}, \ and\ \bibinfo
  {author} {\bibfnamefont {K.}~\bibnamefont {Binder}},\ }\href {\doibase
  10.1039/C6SM00778C} {\bibfield  {journal} {\bibinfo  {journal} {Soft Matter}\
  }\textbf {\bibinfo {volume} {12}},\ \bibinfo {pages} {4944} (\bibinfo {year}
  {2016})}\BibitemShut {NoStop}%
\bibitem [{\citenamefont {Hsu}\ \emph {et~al.}(2010)\citenamefont {Hsu},
  \citenamefont {Paul},\ and\ \citenamefont {Binder}}]{hsu2010standard}%
  \BibitemOpen
  \bibfield  {author} {\bibinfo {author} {\bibfnamefont {H.-P.}\ \bibnamefont
  {Hsu}}, \bibinfo {author} {\bibfnamefont {W.}~\bibnamefont {Paul}}, \ and\
  \bibinfo {author} {\bibfnamefont {K.}~\bibnamefont {Binder}},\ }\href
  {\doibase 10.1021/ma902715e} {\bibfield  {journal} {\bibinfo  {journal}
  {Macromolecules}\ }\textbf {\bibinfo {volume} {43}},\ \bibinfo {pages} {3094}
  (\bibinfo {year} {2010})}\BibitemShut {NoStop}%
\bibitem [{\citenamefont {Salamonczyk}\ \emph {et~al.}(2016)\citenamefont
  {Salamonczyk}, \citenamefont {Zhang}, \citenamefont {Portale}, \citenamefont
  {Zhu}, \citenamefont {Kentzinger}, \citenamefont {Gleeson}, \citenamefont
  {Jakli}, \citenamefont {De~Michele}, \citenamefont {Dhont}, \citenamefont
  {Sprunt},\ and\ \citenamefont {Stiakakis}}]{salamonczyk2016smectic}%
  \BibitemOpen
  \bibfield  {author} {\bibinfo {author} {\bibfnamefont {M.}~\bibnamefont
  {Salamonczyk}}, \bibinfo {author} {\bibfnamefont {J.}~\bibnamefont {Zhang}},
  \bibinfo {author} {\bibfnamefont {G.}~\bibnamefont {Portale}}, \bibinfo
  {author} {\bibfnamefont {C.}~\bibnamefont {Zhu}}, \bibinfo {author}
  {\bibfnamefont {E.}~\bibnamefont {Kentzinger}}, \bibinfo {author}
  {\bibfnamefont {J.~T.}\ \bibnamefont {Gleeson}}, \bibinfo {author}
  {\bibfnamefont {A.}~\bibnamefont {Jakli}}, \bibinfo {author} {\bibfnamefont
  {C.}~\bibnamefont {De~Michele}}, \bibinfo {author} {\bibfnamefont {J.~K.~G.}\
  \bibnamefont {Dhont}}, \bibinfo {author} {\bibfnamefont {S.}~\bibnamefont
  {Sprunt}}, \ and\ \bibinfo {author} {\bibfnamefont {E.}~\bibnamefont
  {Stiakakis}},\ }\href {\doibase 10.1038/ncomms13358} {\bibfield  {journal}
  {\bibinfo  {journal} {Nat. Commun.}\ }\textbf {\bibinfo {volume} {7}},\
  \bibinfo {pages} {13358} (\bibinfo {year} {2016})}\BibitemShut {NoStop}%
\bibitem [{\citenamefont {Barber}\ \emph {et~al.}(1996)\citenamefont {Barber},
  \citenamefont {Dobkin},\ and\ \citenamefont
  {Huhdanpaa}}]{barber1996quickhull}%
  \BibitemOpen
  \bibfield  {author} {\bibinfo {author} {\bibfnamefont {C.~B.}\ \bibnamefont
  {Barber}}, \bibinfo {author} {\bibfnamefont {D.~P.}\ \bibnamefont {Dobkin}},
  \ and\ \bibinfo {author} {\bibfnamefont {H.}~\bibnamefont {Huhdanpaa}},\
  }\href {\doibase 10.1145/235815.235821} {\bibfield  {journal} {\bibinfo
  {journal} {ACM Trans. Math. Softw.}\ }\textbf {\bibinfo {volume} {22}},\
  \bibinfo {pages} {469–483} (\bibinfo {year} {1996})}\BibitemShut {NoStop}%
\bibitem [{\citenamefont {Kantor}\ and\ \citenamefont
  {Nelson}(1987)}]{kantor1987phase}%
  \BibitemOpen
  \bibfield  {author} {\bibinfo {author} {\bibfnamefont {Y.}~\bibnamefont
  {Kantor}}\ and\ \bibinfo {author} {\bibfnamefont {D.~R.}\ \bibnamefont
  {Nelson}},\ }\href {\doibase 10.1103/PhysRevA.36.4020} {\bibfield  {journal}
  {\bibinfo  {journal} {Phys. Rev. A}\ }\textbf {\bibinfo {volume} {36}},\
  \bibinfo {pages} {4020} (\bibinfo {year} {1987})}\BibitemShut {NoStop}%
\bibitem [{\citenamefont {Dahl}\ \emph {et~al.}(2004)\citenamefont {Dahl},
  \citenamefont {Kahn}, \citenamefont {Wilson},\ and\ \citenamefont
  {Discher}}]{dahl2004nuclear}%
  \BibitemOpen
  \bibfield  {author} {\bibinfo {author} {\bibfnamefont {K.~N.}\ \bibnamefont
  {Dahl}}, \bibinfo {author} {\bibfnamefont {S.~M.}\ \bibnamefont {Kahn}},
  \bibinfo {author} {\bibfnamefont {K.~L.}\ \bibnamefont {Wilson}}, \ and\
  \bibinfo {author} {\bibfnamefont {D.~E.}\ \bibnamefont {Discher}},\ }\href
  {https://jcs.biologists.org/content/117/20/4779} {\bibfield  {journal}
  {\bibinfo  {journal} {J. Cell Sci.}\ }\textbf {\bibinfo {volume} {117}},\
  \bibinfo {pages} {4779} (\bibinfo {year} {2004})}\BibitemShut {NoStop}%
\bibitem [{\citenamefont {Frenkel}\ and\ \citenamefont
  {Smit}(2002)}]{frenkel2002understanding}%
  \BibitemOpen
  \bibfield  {author} {\bibinfo {author} {\bibfnamefont {D.}~\bibnamefont
  {Frenkel}}\ and\ \bibinfo {author} {\bibfnamefont {B.}~\bibnamefont {Smit}},\
  }\href@noop {} {\emph {\bibinfo {title} {Understanding Molecular
  Simulation}}},\ \bibinfo {edition} {2nd}\ ed.\ (\bibinfo  {publisher}
  {Academic Press},\ \bibinfo {address} {San Diego, CA},\ \bibinfo {year}
  {2002})\BibitemShut {NoStop}%
\bibitem [{\citenamefont {Glaser}\ \emph {et~al.}(2015)\citenamefont {Glaser},
  \citenamefont {Nguyen}, \citenamefont {Anderson}, \citenamefont {Lui},
  \citenamefont {Spiga}, \citenamefont {Millan}, \citenamefont {Morse},\ and\
  \citenamefont {Glotzer}}]{glaser2015strong}%
  \BibitemOpen
  \bibfield  {author} {\bibinfo {author} {\bibfnamefont {J.}~\bibnamefont
  {Glaser}}, \bibinfo {author} {\bibfnamefont {T.~D.}\ \bibnamefont {Nguyen}},
  \bibinfo {author} {\bibfnamefont {J.~A.}\ \bibnamefont {Anderson}}, \bibinfo
  {author} {\bibfnamefont {P.}~\bibnamefont {Lui}}, \bibinfo {author}
  {\bibfnamefont {F.}~\bibnamefont {Spiga}}, \bibinfo {author} {\bibfnamefont
  {J.~A.}\ \bibnamefont {Millan}}, \bibinfo {author} {\bibfnamefont {D.~C.}\
  \bibnamefont {Morse}}, \ and\ \bibinfo {author} {\bibfnamefont {S.~C.}\
  \bibnamefont {Glotzer}},\ }\href {\doibase
  https://doi.org/10.1016/j.cpc.2015.02.028} {\bibfield  {journal} {\bibinfo
  {journal} {Comput. Phys. Commun.}\ }\textbf {\bibinfo {volume} {192}},\
  \bibinfo {pages} {97} (\bibinfo {year} {2015})}\BibitemShut {NoStop}%
\bibitem [{\citenamefont {Anderson}\ \emph {et~al.}(2020)\citenamefont
  {Anderson}, \citenamefont {Glaser},\ and\ \citenamefont
  {Glotzer}}]{anderson2020hoomd}%
  \BibitemOpen
  \bibfield  {author} {\bibinfo {author} {\bibfnamefont {J.~A.}\ \bibnamefont
  {Anderson}}, \bibinfo {author} {\bibfnamefont {J.}~\bibnamefont {Glaser}}, \
  and\ \bibinfo {author} {\bibfnamefont {S.~C.}\ \bibnamefont {Glotzer}},\
  }\href {\doibase https://doi.org/10.1016/j.commatsci.2019.109363} {\bibfield
  {journal} {\bibinfo  {journal} {Comput. Mater. Sci.}\ }\textbf {\bibinfo
  {volume} {173}},\ \bibinfo {pages} {109363} (\bibinfo {year}
  {2020})}\BibitemShut {NoStop}%
\bibitem [{\citenamefont {Howard}\ \emph {et~al.}(2016)\citenamefont {Howard},
  \citenamefont {Anderson}, \citenamefont {Nikoubashman}, \citenamefont
  {Glotzer},\ and\ \citenamefont {Panagiotopoulos}}]{howard2016efficient}%
  \BibitemOpen
  \bibfield  {author} {\bibinfo {author} {\bibfnamefont {M.~P.}\ \bibnamefont
  {Howard}}, \bibinfo {author} {\bibfnamefont {J.~A.}\ \bibnamefont
  {Anderson}}, \bibinfo {author} {\bibfnamefont {A.}~\bibnamefont
  {Nikoubashman}}, \bibinfo {author} {\bibfnamefont {S.~C.}\ \bibnamefont
  {Glotzer}}, \ and\ \bibinfo {author} {\bibfnamefont {A.~Z.}\ \bibnamefont
  {Panagiotopoulos}},\ }\href {\doibase
  https://doi.org/10.1016/j.cpc.2016.02.003} {\bibfield  {journal} {\bibinfo
  {journal} {Comput. Phys. Commun.}\ }\textbf {\bibinfo {volume} {203}},\
  \bibinfo {pages} {45} (\bibinfo {year} {2016})}\BibitemShut {NoStop}%
\bibitem [{\citenamefont {Zhang}\ and\ \citenamefont
  {Chen}(2001)}]{zhang2001efficient}%
  \BibitemOpen
  \bibfield  {author} {\bibinfo {author} {\bibfnamefont {C.}~\bibnamefont
  {Zhang}}\ and\ \bibinfo {author} {\bibfnamefont {T.}~\bibnamefont {Chen}},\
  }in\ \href {\doibase 10.1109/ICIP.2001.958278} {\emph {\bibinfo {booktitle}
  {IEEE Image Proc.}}},\ Vol.~\bibinfo {volume} {3}\ (\bibinfo {year} {2001})\
  pp.\ \bibinfo {pages} {935--938}\BibitemShut {NoStop}%
\bibitem [{\citenamefont {Trukhina}\ and\ \citenamefont
  {Schilling}(2008)}]{trukhina2008computer}%
  \BibitemOpen
  \bibfield  {author} {\bibinfo {author} {\bibfnamefont {Y.}~\bibnamefont
  {Trukhina}}\ and\ \bibinfo {author} {\bibfnamefont {T.}~\bibnamefont
  {Schilling}},\ }\href {\doibase 10.1103/PhysRevE.77.011701} {\bibfield
  {journal} {\bibinfo  {journal} {Phys. Rev. E}\ }\textbf {\bibinfo {volume}
  {77}},\ \bibinfo {pages} {011701} (\bibinfo {year} {2008})}\BibitemShut
  {NoStop}%
\bibitem [{\citenamefont {Kos}\ and\ \citenamefont
  {Ravnik}(2016)}]{kos2016relevance}%
  \BibitemOpen
  \bibfield  {author} {\bibinfo {author} {\bibfnamefont {{\v Z}.}~\bibnamefont
  {Kos}}\ and\ \bibinfo {author} {\bibfnamefont {M.}~\bibnamefont {Ravnik}},\
  }\href {\doibase 10.1039/C5SM02417J} {\bibfield  {journal} {\bibinfo
  {journal} {Soft Matter}\ }\textbf {\bibinfo {volume} {12}},\ \bibinfo {pages}
  {1313} (\bibinfo {year} {2016})}\BibitemShut {NoStop}%
\bibitem [{\citenamefont {Williams}(1986)}]{williams1986two}%
  \BibitemOpen
  \bibfield  {author} {\bibinfo {author} {\bibfnamefont {R.~D.}\ \bibnamefont
  {Williams}},\ }\href {\doibase 10.1088/0305-4470/19/16/019} {\bibfield
  {journal} {\bibinfo  {journal} {J. Phys. A}\ }\textbf {\bibinfo {volume}
  {19}},\ \bibinfo {pages} {3211} (\bibinfo {year} {1986})}\BibitemShut
  {NoStop}%
\bibitem [{\citenamefont {{Terentjev, E. M.}}(1995)}]{terentjev1995density}%
  \BibitemOpen
  \bibfield  {author} {\bibinfo {author} {\bibnamefont {{Terentjev, E. M.}}},\
  }\href {\doibase 10.1051/jp2:1995120} {\bibfield  {journal} {\bibinfo
  {journal} {J. Phys. II France}\ }\textbf {\bibinfo {volume} {5}},\ \bibinfo
  {pages} {159} (\bibinfo {year} {1995})}\BibitemShut {NoStop}%
\bibitem [{\citenamefont {Tortora}\ and\ \citenamefont
  {Doye}(2017{\natexlab{a}})}]{tortora2017perturbative}%
  \BibitemOpen
  \bibfield  {author} {\bibinfo {author} {\bibfnamefont {M.~M.~C.}\
  \bibnamefont {Tortora}}\ and\ \bibinfo {author} {\bibfnamefont {J.~P.~K.}\
  \bibnamefont {Doye}},\ }\href {\doibase 10.1063/1.4982934} {\bibfield
  {journal} {\bibinfo  {journal} {J. Chem. Phys.}\ }\textbf {\bibinfo {volume}
  {146}},\ \bibinfo {pages} {184504} (\bibinfo {year}
  {2017}{\natexlab{a}})}\BibitemShut {NoStop}%
\bibitem [{\citenamefont {Woodward}(1991)}]{woodward1991density}%
  \BibitemOpen
  \bibfield  {author} {\bibinfo {author} {\bibfnamefont {C.~E.}\ \bibnamefont
  {Woodward}},\ }\href {\doibase 10.1063/1.459787} {\bibfield  {journal}
  {\bibinfo  {journal} {J. Chem. Phys.}\ }\textbf {\bibinfo {volume} {94}},\
  \bibinfo {pages} {3183} (\bibinfo {year} {1991})}\BibitemShut {NoStop}%
\bibitem [{\citenamefont {Fynewever}\ and\ \citenamefont
  {Yethiraj}(1998)}]{fynewever1998phase}%
  \BibitemOpen
  \bibfield  {author} {\bibinfo {author} {\bibfnamefont {H.}~\bibnamefont
  {Fynewever}}\ and\ \bibinfo {author} {\bibfnamefont {A.}~\bibnamefont
  {Yethiraj}},\ }\href {\doibase 10.1063/1.475534} {\bibfield  {journal}
  {\bibinfo  {journal} {J. Chem. Phys.}\ }\textbf {\bibinfo {volume} {108}},\
  \bibinfo {pages} {1636} (\bibinfo {year} {1998})}\BibitemShut {NoStop}%
\bibitem [{\citenamefont {Parsons}(1979)}]{parsons1979nematic}%
  \BibitemOpen
  \bibfield  {author} {\bibinfo {author} {\bibfnamefont {J.~D.}\ \bibnamefont
  {Parsons}},\ }\href {\doibase 10.1103/PhysRevA.19.1225} {\bibfield  {journal}
  {\bibinfo  {journal} {Phys. Rev. A}\ }\textbf {\bibinfo {volume} {19}},\
  \bibinfo {pages} {1225} (\bibinfo {year} {1979})}\BibitemShut {NoStop}%
\bibitem [{\citenamefont {Lee}(1987)}]{lee1987numerical}%
  \BibitemOpen
  \bibfield  {author} {\bibinfo {author} {\bibfnamefont {S.~D.}\ \bibnamefont
  {Lee}},\ }\href {\doibase 10.1063/1.452811} {\bibfield  {journal} {\bibinfo
  {journal} {J. Chem. Phys.}\ }\textbf {\bibinfo {volume} {87}},\ \bibinfo
  {pages} {4972} (\bibinfo {year} {1987})}\BibitemShut {NoStop}%
\bibitem [{\citenamefont {Tortora}\ and\ \citenamefont
  {Doye}(2018)}]{tortora2018incorporating}%
  \BibitemOpen
  \bibfield  {author} {\bibinfo {author} {\bibfnamefont {M.~M.~C.}\
  \bibnamefont {Tortora}}\ and\ \bibinfo {author} {\bibfnamefont {J.~P.~K.}\
  \bibnamefont {Doye}},\ }\href {\doibase 10.1080/00268976.2018.1464226}
  {\bibfield  {journal} {\bibinfo  {journal} {Mol. Phys.}\ }\textbf {\bibinfo
  {volume} {116}},\ \bibinfo {pages} {2773} (\bibinfo {year}
  {2018})}\BibitemShut {NoStop}%
\bibitem [{\citenamefont {Milchev}\ \emph
  {et~al.}(2018{\natexlab{b}})\citenamefont {Milchev}, \citenamefont {Egorov},
  \citenamefont {Binder},\ and\ \citenamefont
  {Nikoubashman}}]{milchev2018nematic}%
  \BibitemOpen
  \bibfield  {author} {\bibinfo {author} {\bibfnamefont {A.}~\bibnamefont
  {Milchev}}, \bibinfo {author} {\bibfnamefont {S.~A.}\ \bibnamefont {Egorov}},
  \bibinfo {author} {\bibfnamefont {K.}~\bibnamefont {Binder}}, \ and\ \bibinfo
  {author} {\bibfnamefont {A.}~\bibnamefont {Nikoubashman}},\ }\href {\doibase
  10.1063/1.5049630} {\bibfield  {journal} {\bibinfo  {journal} {J. Chem.
  Phys.}\ }\textbf {\bibinfo {volume} {149}},\ \bibinfo {pages} {174909}
  (\bibinfo {year} {2018}{\natexlab{b}})}\BibitemShut {NoStop}%
\bibitem [{\citenamefont {Tortora}\ and\ \citenamefont
  {Doye}(2017{\natexlab{b}})}]{tortora2017hierarchical}%
  \BibitemOpen
  \bibfield  {author} {\bibinfo {author} {\bibfnamefont {M.~M.~C.}\
  \bibnamefont {Tortora}}\ and\ \bibinfo {author} {\bibfnamefont {J.~P.~K.}\
  \bibnamefont {Doye}},\ }\href {\doibase 10.1063/1.5002666} {\bibfield
  {journal} {\bibinfo  {journal} {J. Chem. Phys.}\ }\textbf {\bibinfo {volume}
  {147}},\ \bibinfo {pages} {224504} (\bibinfo {year}
  {2017}{\natexlab{b}})}\BibitemShut {NoStop}%
\bibitem [{\citenamefont {Tortora}\ \emph {et~al.}(2020)\citenamefont
  {Tortora}, \citenamefont {Mishra}, \citenamefont {Pre{\v s}ern},\ and\
  \citenamefont {Doye}}]{tortora2020chiral}%
  \BibitemOpen
  \bibfield  {author} {\bibinfo {author} {\bibfnamefont {M.~M.~C.}\
  \bibnamefont {Tortora}}, \bibinfo {author} {\bibfnamefont {G.}~\bibnamefont
  {Mishra}}, \bibinfo {author} {\bibfnamefont {D.}~\bibnamefont {Pre{\v
  s}ern}}, \ and\ \bibinfo {author} {\bibfnamefont {J.~P.~K.}\ \bibnamefont
  {Doye}},\ }\href {\doibase 10.1126/sciadv.aaw8331} {\bibfield  {journal}
  {\bibinfo  {journal} {Sci. Adv.}\ }\textbf {\bibinfo {volume} {6}} (\bibinfo
  {year} {2020}),\ 10.1126/sciadv.aaw8331}\BibitemShut {NoStop}%
\bibitem [{\citenamefont {Yokoyama}(1997)}]{yokoyama1997density}%
  \BibitemOpen
  \bibfield  {author} {\bibinfo {author} {\bibfnamefont {H.}~\bibnamefont
  {Yokoyama}},\ }\href {\doibase 10.1103/PhysRevE.55.2938} {\bibfield
  {journal} {\bibinfo  {journal} {Phys. Rev. E}\ }\textbf {\bibinfo {volume}
  {55}},\ \bibinfo {pages} {2938} (\bibinfo {year} {1997})}\BibitemShut
  {NoStop}%
\bibitem [{\citenamefont {Meyer}(1982)}]{meyer1982macroscopic}%
  \BibitemOpen
  \bibfield  {author} {\bibinfo {author} {\bibfnamefont {R.~B.}\ \bibnamefont
  {Meyer}},\ }in\ \href {\doibase
  https://doi.org/10.1016/B978-0-12-174680-3.50011-8} {\emph {\bibinfo
  {booktitle} {Polymer Liquid Crystals}}},\ \bibinfo {editor} {edited by\
  \bibinfo {editor} {\bibfnamefont {A.}~\bibnamefont {Ciferri}}, \bibinfo
  {editor} {\bibfnamefont {W.}~\bibnamefont {Krigbaum}}, \ and\ \bibinfo
  {editor} {\bibfnamefont {R.~B.}\ \bibnamefont {Meyer}}}\ (\bibinfo
  {publisher} {Academic Press},\ \bibinfo {year} {1982})\ pp.\ \bibinfo {pages}
  {133--163}\BibitemShut {NoStop}%
\bibitem [{\citenamefont {Vitelli}\ and\ \citenamefont
  {Nelson}(2006)}]{vitelli2006nematic}%
  \BibitemOpen
  \bibfield  {author} {\bibinfo {author} {\bibfnamefont {V.}~\bibnamefont
  {Vitelli}}\ and\ \bibinfo {author} {\bibfnamefont {D.~R.}\ \bibnamefont
  {Nelson}},\ }\href {\doibase 10.1103/PhysRevE.74.021711} {\bibfield
  {journal} {\bibinfo  {journal} {Phys. Rev. E}\ }\textbf {\bibinfo {volume}
  {74}},\ \bibinfo {pages} {021711} (\bibinfo {year} {2006})}\BibitemShut
  {NoStop}%
\bibitem [{\citenamefont {van Roij}\ \emph {et~al.}(2000)\citenamefont {van
  Roij}, \citenamefont {Dijkstra},\ and\ \citenamefont
  {Evans}}]{vanRoij2000orientational}%
  \BibitemOpen
  \bibfield  {author} {\bibinfo {author} {\bibfnamefont {R.}~\bibnamefont {van
  Roij}}, \bibinfo {author} {\bibfnamefont {M.}~\bibnamefont {Dijkstra}}, \
  and\ \bibinfo {author} {\bibfnamefont {R.}~\bibnamefont {Evans}},\ }\href
  {\doibase 10.1209/epl/i2000-00155-0} {\bibfield  {journal} {\bibinfo
  {journal} {EPL}\ }\textbf {\bibinfo {volume} {49}},\ \bibinfo {pages} {350}
  (\bibinfo {year} {2000})}\BibitemShut {NoStop}%
\bibitem [{\citenamefont {Ivanov}\ \emph {et~al.}(2013)\citenamefont {Ivanov},
  \citenamefont {Rodionova}, \citenamefont {Martemyanova}, \citenamefont
  {Stukan}, \citenamefont {Müller}, \citenamefont {Paul},\ and\ \citenamefont
  {Binder}}]{ivanov2013wall}%
  \BibitemOpen
  \bibfield  {author} {\bibinfo {author} {\bibfnamefont {V.~A.}\ \bibnamefont
  {Ivanov}}, \bibinfo {author} {\bibfnamefont {A.~S.}\ \bibnamefont
  {Rodionova}}, \bibinfo {author} {\bibfnamefont {J.~A.}\ \bibnamefont
  {Martemyanova}}, \bibinfo {author} {\bibfnamefont {M.~R.}\ \bibnamefont
  {Stukan}}, \bibinfo {author} {\bibfnamefont {M.}~\bibnamefont {Müller}},
  \bibinfo {author} {\bibfnamefont {W.}~\bibnamefont {Paul}}, \ and\ \bibinfo
  {author} {\bibfnamefont {K.}~\bibnamefont {Binder}},\ }\href {\doibase
  10.1063/1.4810745} {\bibfield  {journal} {\bibinfo  {journal} {J. Chem.
  Phys.}\ }\textbf {\bibinfo {volume} {138}},\ \bibinfo {pages} {234903}
  (\bibinfo {year} {2013})}\BibitemShut {NoStop}%
\bibitem [{\citenamefont {Nelson}(2002)}]{nelson2002toward}%
  \BibitemOpen
  \bibfield  {author} {\bibinfo {author} {\bibfnamefont {D.~R.}\ \bibnamefont
  {Nelson}},\ }\href {\doibase 10.1021/nl0202096} {\bibfield  {journal}
  {\bibinfo  {journal} {Nano Lett.}\ }\textbf {\bibinfo {volume} {2}},\
  \bibinfo {pages} {1125} (\bibinfo {year} {2002})}\BibitemShut {NoStop}%
\bibitem [{\citenamefont {{T.C. Lubensky}}\ and\ \citenamefont {{J
  Prost}}(1992)}]{lubensky1992orientational}%
  \BibitemOpen
  \bibfield  {author} {\bibinfo {author} {\bibnamefont {{T.C. Lubensky}}}\ and\
  \bibinfo {author} {\bibnamefont {{J Prost}}},\ }\href {\doibase
  10.1051/jp2:1992133} {\bibfield  {journal} {\bibinfo  {journal} {J. Phys. II
  France}\ }\textbf {\bibinfo {volume} {2}},\ \bibinfo {pages} {371} (\bibinfo
  {year} {1992})}\BibitemShut {NoStop}%
\bibitem [{\citenamefont {Banerjee}\ \emph {et~al.}(2005)\citenamefont
  {Banerjee}, \citenamefont {Dhillon}, \citenamefont {Ghosh},\ and\
  \citenamefont {Sra}}]{banerjee2005clustering}%
  \BibitemOpen
  \bibfield  {author} {\bibinfo {author} {\bibfnamefont {A.}~\bibnamefont
  {Banerjee}}, \bibinfo {author} {\bibfnamefont {I.~S.}\ \bibnamefont
  {Dhillon}}, \bibinfo {author} {\bibfnamefont {J.}~\bibnamefont {Ghosh}}, \
  and\ \bibinfo {author} {\bibfnamefont {S.}~\bibnamefont {Sra}},\ }\href
  {https://www.jmlr.org/papers/volume6/banerjee05a/banerjee05a.pdf} {\bibfield
  {journal} {\bibinfo  {journal} {J. Mach. Learn. Res.}\ }\textbf {\bibinfo
  {volume} {6}},\ \bibinfo {pages} {1345–1382} (\bibinfo {year}
  {2005})}\BibitemShut {NoStop}%
\bibitem [{\citenamefont {{Robert}}\ \emph {et~al.}(1998)\citenamefont
  {{Robert}}, \citenamefont {{Roux}}, \citenamefont {{Harvey}}, \citenamefont
  {{Dunlop}}, \citenamefont {{Daly}},\ and\ \citenamefont
  {{Glassmeier}}}]{robert1998tetrahedron}%
  \BibitemOpen
  \bibfield  {author} {\bibinfo {author} {\bibfnamefont {P.}~\bibnamefont
  {{Robert}}}, \bibinfo {author} {\bibfnamefont {A.}~\bibnamefont {{Roux}}},
  \bibinfo {author} {\bibfnamefont {C.~C.}\ \bibnamefont {{Harvey}}}, \bibinfo
  {author} {\bibfnamefont {M.~W.}\ \bibnamefont {{Dunlop}}}, \bibinfo {author}
  {\bibfnamefont {P.~W.}\ \bibnamefont {{Daly}}}, \ and\ \bibinfo {author}
  {\bibfnamefont {K.-H.}\ \bibnamefont {{Glassmeier}}},\ }\href
  {http://www.issibern.ch/forads/sr-001-13.pdf} {\bibfield  {journal} {\bibinfo
   {journal} {ISSI Sci. Rep.}\ }\textbf {\bibinfo {volume} {1}},\ \bibinfo
  {pages} {323} (\bibinfo {year} {1998})}\BibitemShut {NoStop}%
\bibitem [{\citenamefont {de~Braaf}\ \emph {et~al.}(2017)\citenamefont
  {de~Braaf}, \citenamefont {Oshima~Menegon}, \citenamefont {Paquay},\ and\
  \citenamefont {van~der Schoot}}]{debraaf2017self}%
  \BibitemOpen
  \bibfield  {author} {\bibinfo {author} {\bibfnamefont {B.}~\bibnamefont
  {de~Braaf}}, \bibinfo {author} {\bibfnamefont {M.}~\bibnamefont
  {Oshima~Menegon}}, \bibinfo {author} {\bibfnamefont {S.}~\bibnamefont
  {Paquay}}, \ and\ \bibinfo {author} {\bibfnamefont {P.}~\bibnamefont {van~der
  Schoot}},\ }\href {\doibase 10.1063/1.5000228} {\bibfield  {journal}
  {\bibinfo  {journal} {J. Chem. Phys.}\ }\textbf {\bibinfo {volume} {147}},\
  \bibinfo {pages} {244901} (\bibinfo {year} {2017})}\BibitemShut {NoStop}%
\bibitem [{\citenamefont {Shin}\ \emph {et~al.}(2008)\citenamefont {Shin},
  \citenamefont {Bowick},\ and\ \citenamefont {Xing}}]{shin2008topological}%
  \BibitemOpen
  \bibfield  {author} {\bibinfo {author} {\bibfnamefont {H.}~\bibnamefont
  {Shin}}, \bibinfo {author} {\bibfnamefont {M.~J.}\ \bibnamefont {Bowick}}, \
  and\ \bibinfo {author} {\bibfnamefont {X.}~\bibnamefont {Xing}},\ }\href
  {\doibase 10.1103/PhysRevLett.101.037802} {\bibfield  {journal} {\bibinfo
  {journal} {Phys. Rev. Lett.}\ }\textbf {\bibinfo {volume} {101}},\ \bibinfo
  {pages} {037802} (\bibinfo {year} {2008})}\BibitemShut {NoStop}%
\bibitem [{\citenamefont {Fern\'andez-Nieves}\ \emph
  {et~al.}(2007)\citenamefont {Fern\'andez-Nieves}, \citenamefont {Link},
  \citenamefont {M\'arquez},\ and\ \citenamefont
  {Weitz}}]{fernandez2007topological}%
  \BibitemOpen
  \bibfield  {author} {\bibinfo {author} {\bibfnamefont {A.}~\bibnamefont
  {Fern\'andez-Nieves}}, \bibinfo {author} {\bibfnamefont {D.~R.}\ \bibnamefont
  {Link}}, \bibinfo {author} {\bibfnamefont {M.}~\bibnamefont {M\'arquez}}, \
  and\ \bibinfo {author} {\bibfnamefont {D.~A.}\ \bibnamefont {Weitz}},\ }\href
  {\doibase 10.1103/PhysRevLett.98.087801} {\bibfield  {journal} {\bibinfo
  {journal} {Phys. Rev. Lett.}\ }\textbf {\bibinfo {volume} {98}},\ \bibinfo
  {pages} {087801} (\bibinfo {year} {2007})}\BibitemShut {NoStop}%
\bibitem [{\citenamefont {Groh}\ and\ \citenamefont
  {Dietrich}(1999)}]{groh1999fluids}%
  \BibitemOpen
  \bibfield  {author} {\bibinfo {author} {\bibfnamefont {B.}~\bibnamefont
  {Groh}}\ and\ \bibinfo {author} {\bibfnamefont {S.}~\bibnamefont
  {Dietrich}},\ }\href {\doibase 10.1103/PhysRevE.59.4216} {\bibfield
  {journal} {\bibinfo  {journal} {Phys. Rev. E}\ }\textbf {\bibinfo {volume}
  {59}},\ \bibinfo {pages} {4216} (\bibinfo {year} {1999})}\BibitemShut
  {NoStop}%
\bibitem [{\citenamefont {Milchev}\ \emph {et~al.}(2021)\citenamefont
  {Milchev}, \citenamefont {Egorov},\ and\ \citenamefont
  {Binder}}]{milchev2021phase}%
  \BibitemOpen
  \bibfield  {author} {\bibinfo {author} {\bibfnamefont {A.}~\bibnamefont
  {Milchev}}, \bibinfo {author} {\bibfnamefont {S.~A.}\ \bibnamefont {Egorov}},
  \ and\ \bibinfo {author} {\bibfnamefont {K.}~\bibnamefont {Binder}},\ }\href
  {\doibase 10.1021/acs.macromol.1c00785} {\bibfield  {journal} {\bibinfo
  {journal} {Macromolecules}\ }\textbf {\bibinfo {volume} {54}},\ \bibinfo
  {pages} {6312} (\bibinfo {year} {2021})}\BibitemShut {NoStop}%
\bibitem [{\citenamefont {Petrov}\ and\ \citenamefont
  {Harvey}(2008)}]{petrov2008packaging}%
  \BibitemOpen
  \bibfield  {author} {\bibinfo {author} {\bibfnamefont {A.~S.}\ \bibnamefont
  {Petrov}}\ and\ \bibinfo {author} {\bibfnamefont {S.~C.}\ \bibnamefont
  {Harvey}},\ }\href {\doibase 10.1529/biophysj.108.131797} {\bibfield
  {journal} {\bibinfo  {journal} {Biophys. J.}\ }\textbf {\bibinfo {volume}
  {95}},\ \bibinfo {pages} {497} (\bibinfo {year} {2008})}\BibitemShut
  {NoStop}%
\bibitem [{\citenamefont {Wang}\ \emph {et~al.}(2016)\citenamefont {Wang},
  \citenamefont {Miller}, \citenamefont {Bukusoglu}, \citenamefont {de~Pablo},\
  and\ \citenamefont {Abbott}}]{wang2016topological}%
  \BibitemOpen
  \bibfield  {author} {\bibinfo {author} {\bibfnamefont {X.}~\bibnamefont
  {Wang}}, \bibinfo {author} {\bibfnamefont {D.~S.}\ \bibnamefont {Miller}},
  \bibinfo {author} {\bibfnamefont {E.}~\bibnamefont {Bukusoglu}}, \bibinfo
  {author} {\bibfnamefont {J.~J.}\ \bibnamefont {de~Pablo}}, \ and\ \bibinfo
  {author} {\bibfnamefont {N.~L.}\ \bibnamefont {Abbott}},\ }\href {\doibase
  10.1038/nmat4421} {\bibfield  {journal} {\bibinfo  {journal} {Nat. Mater.}\
  }\textbf {\bibinfo {volume} {15}},\ \bibinfo {pages} {106} (\bibinfo {year}
  {2016})}\BibitemShut {NoStop}%
\bibitem [{\citenamefont {Leforestier}\ and\ \citenamefont
  {Livolant}(2010)}]{leforestier2010bacteriophage}%
  \BibitemOpen
  \bibfield  {author} {\bibinfo {author} {\bibfnamefont {A.}~\bibnamefont
  {Leforestier}}\ and\ \bibinfo {author} {\bibfnamefont {F.}~\bibnamefont
  {Livolant}},\ }\href {\doibase https://doi.org/10.1016/j.jmb.2009.11.047}
  {\bibfield  {journal} {\bibinfo  {journal} {J. Mol. Biol.}\ }\textbf
  {\bibinfo {volume} {396}},\ \bibinfo {pages} {384} (\bibinfo {year}
  {2010})}\BibitemShut {NoStop}%
\bibitem [{\citenamefont {Marenduzzo}\ \emph {et~al.}(2009)\citenamefont
  {Marenduzzo}, \citenamefont {Orlandini}, \citenamefont {Stasiak},
  \citenamefont {Sumners}, \citenamefont {Tubiana},\ and\ \citenamefont
  {Micheletti}}]{marenduzzo2009dna}%
  \BibitemOpen
  \bibfield  {author} {\bibinfo {author} {\bibfnamefont {D.}~\bibnamefont
  {Marenduzzo}}, \bibinfo {author} {\bibfnamefont {E.}~\bibnamefont
  {Orlandini}}, \bibinfo {author} {\bibfnamefont {A.}~\bibnamefont {Stasiak}},
  \bibinfo {author} {\bibfnamefont {D.~W.}\ \bibnamefont {Sumners}}, \bibinfo
  {author} {\bibfnamefont {L.}~\bibnamefont {Tubiana}}, \ and\ \bibinfo
  {author} {\bibfnamefont {C.}~\bibnamefont {Micheletti}},\ }\href {\doibase
  10.1073/pnas.0907524106} {\bibfield  {journal} {\bibinfo  {journal} {Proc.
  Natl. Acad. Sci. U.S.A.}\ }\textbf {\bibinfo {volume} {106}},\ \bibinfo
  {pages} {22269} (\bibinfo {year} {2009})}\BibitemShut {NoStop}%
\end{thebibliography}
\end{document}

%----------------------------------------------------------------------------------------
%	TITLE SECTION
%----------------------------------------------------------------------------------------

\title{Supplementary information: Orientational wetting and topological transitions in confined solutions of semi-flexible polymers}
\date{\today}
\author{Maxime M. C. Tortora}
\thanks{Correspondence to \href{mailto:maxime.tortora@ens-lyon.fr}{maxime.tortora@ens-lyon.fr}}
\author{Daniel Jost}
\affiliation{Universit\'e de Lyon, ENS de Lyon, Univ Claude Bernard, CNRS, Laboratoire de Biologie et Mod\'elisation de la Cellule, Lyon, France}

%----------------------------------------------------------------------------------------

%\pacs{23.23.+x, 56.65.Dy}
%\keywords{Density functional theory, liquid crystals, multi-scale modelling, chirality, perturbation theory.}

\maketitle % Insert title

%----------------------------------------------------------------------------------------
%	ARTICLE CONTENTS
%----------------------------------------------------------------------------------------

\begin{figure}[htpb]
  \includegraphics[width=\columnwidth]{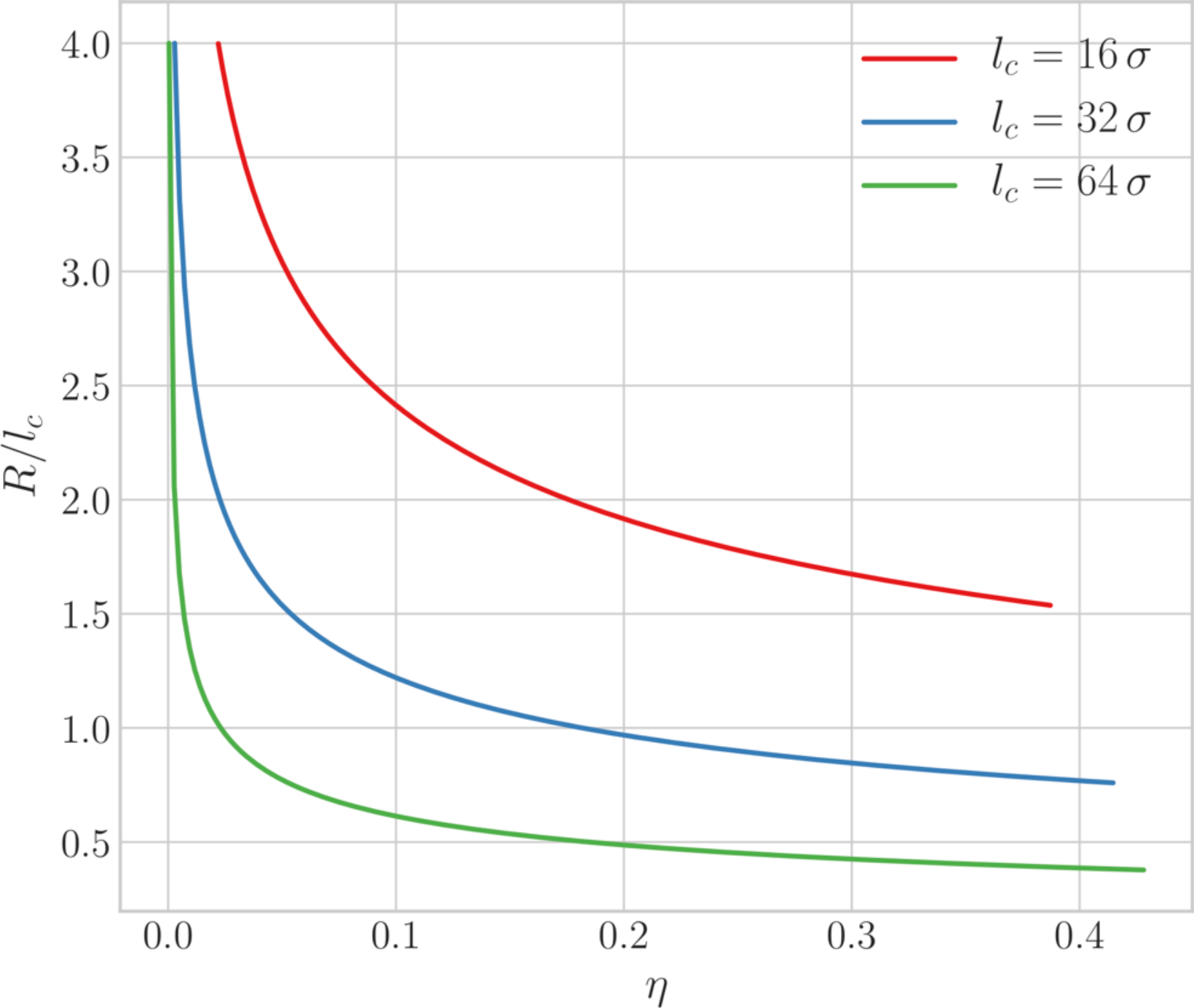}
  \caption{\label{figS1}Mean membrane radius ($R$) as a function of polymer volume fraction ($\eta$) for the different systems investigated in the main text.}
\end{figure}

\section{Free energy calculations} \label{sec:free}

The density variations $\Delta \mathscr{F}$ of the equilibrium free energy may be evaluated using Eqs.~(9)--(12) of the main text,
\begin{equation}
  \label{eq:deltaf_th}
  \Delta \mathscr{F}(\eta) = \mathscr{F}_0\big[\psi_\eta^{\rm eq}\big]-\mathscr{F}_0\big[\psi_{\eta_0}^{\rm eq}\big],
\end{equation}
with $\eta_0$ the polymer density in the simulation initial state. Eq.~\eqref{eq:deltaf_th} may be directly compared against the free energy computed by thermodynamic integration along the compression path of our simulations,
\begin{equation}
  \label{eq:deltaf_sim}
  \Delta \mathscr{F}(\eta) = -\int_{V_0}^V dV'\, \Pi = \int_{\eta_0}^\eta V'\frac{d\eta'}{\eta'} \,\Pi(\eta'),
\end{equation}
where the osmotic pressure $\Pi = \Tr{\mathcal{P}}/3$ of the encapsulated polymer solution was obtained from the classical pressure tensor $\mathcal{P}$ as given by the virial theorem~\cite{gray1984theory},
\begin{equation*}
  \mathcal{P}^{\alpha\beta} = \rho k_B T\, \delta_{\alpha\beta}+ \frac{1}{3V}\Bigg\langle\sum_{k} r_{k}^\alpha F_{k}^\beta\Bigg\rangle,
\end{equation*}
with $\rho\equiv N/V=\eta N_m/v_c$ the monomer density and $\mathbf{F}_{k}$ the total instantaneous force exerted on a monomer $k$.

\section{Averaged normal tensor from bond vector components} \label{sec:Q_tens}

\begin{figure*}[htpb]
  \includegraphics[width=1.5\columnwidth]{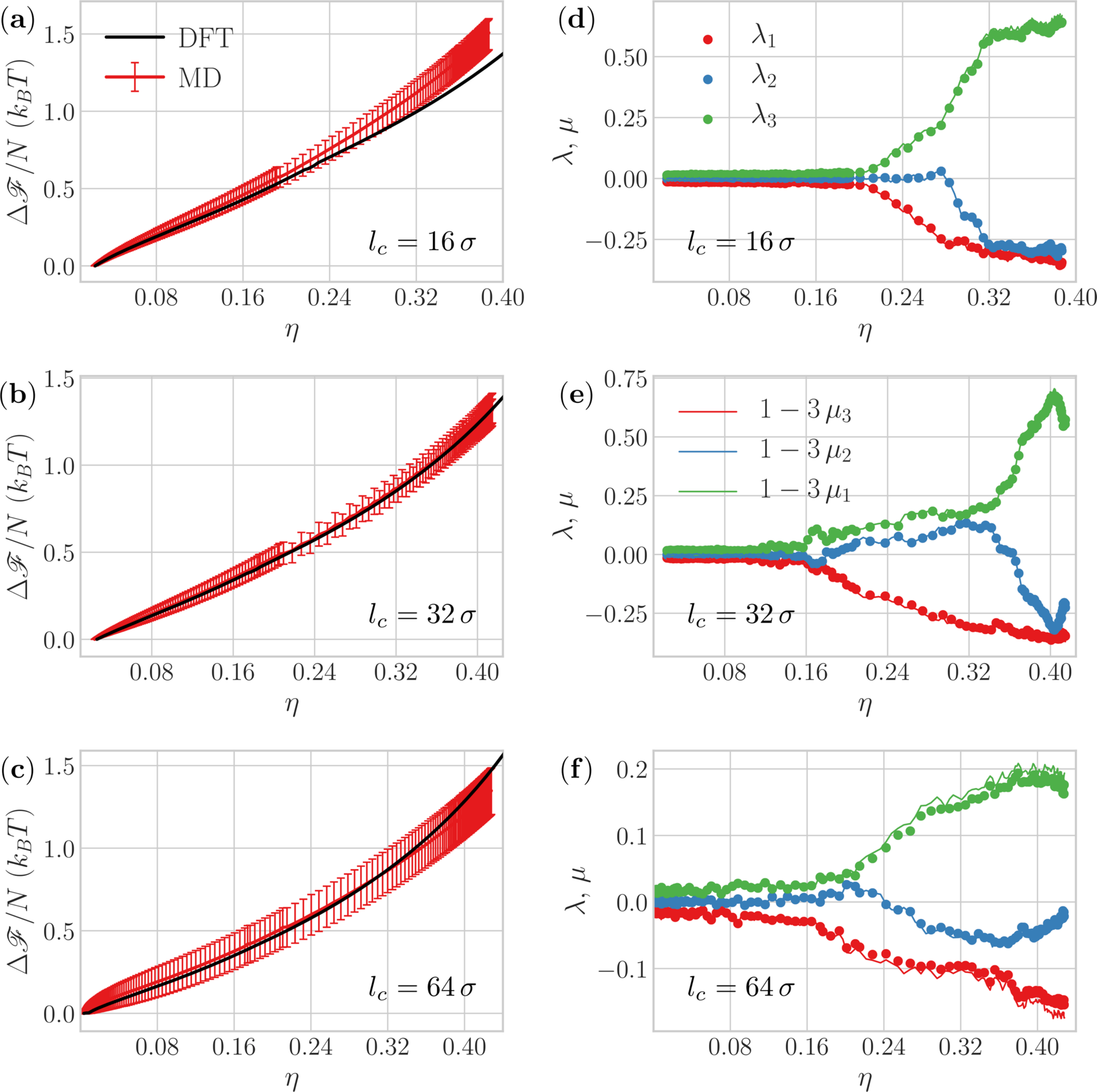}
  \caption{\label{figS2}Equilibrium free energy and shape fluctuations of confined DNA-like polymers. (a)-(c) Density variations of the free energy ($\Delta \mathscr{F}$) of the chains as a function of polymer volume fraction ($\eta$) for the various contour lengths $l_c$ considered in the main text. DFT predictions are derived by numerical resolution of Eq.~\eqref{eq:deltaf_th}, and MD results are computed from the simulations by thermodynamic integration of Eq.~\eqref{eq:deltaf_sim} through standard quadrature methods. Error bars are obtained as described in the main text. (d)-(f) Nematic $\mathcal{Q}$- ($\lambda_i$) and normal $\mathcal{N}$-tensor eigenvalues ($\mu_i$) as a function of $\eta$. Note the close agreement with the theoretical relation Eq.~\eqref{eq:eig_perp} expected from the equipartition theorem.}
\end{figure*}

\begin{lemma}
\label{lemma:dyadic}
Let $\mathbf{v} \equiv \begin{bmatrix} v_x\; v_y \;v_z \end{bmatrix}$ be an arbitrary unit vector, and $\begin{bmatrix} \mathbf{v} \end{bmatrix}_\times$ the skew-symmetric matrix such that
\begin{gather*}
   \mathbf{v} \times \mathbf{x} = \begin{bmatrix} \mathbf{v} \end{bmatrix}_\times \cdot \mathbf{x} \qquad \forall \mathbf{x} \in \mathbb{R}^3 \\ \big \Updownarrow \\ \begin{bmatrix} \mathbf{v} \end{bmatrix}_\times = 
  \begin{bmatrix} 
  0 & -v_z & v_y \\
  v_z & 0 & -v_x \\
  -v_y & v_x & 0
  \end{bmatrix},
\end{gather*}
Then,
\begin{equation}
  \label{eq:dyadic_cross}
  \mathbf{v} \otimes \mathbf{v} = \begin{bmatrix} \mathbf{v} \end{bmatrix}_\times^2 + \mathcal{I},
\end{equation}
with $\mathcal{I}$ the 3D identity matrix.
\end{lemma}

\begin{proof}
By definition of the dyadic product,
\begin{equation*}
    \mathbf{v} \otimes \mathbf{v} \equiv \begin{bmatrix} v_i v_j  \end{bmatrix} = \mathbf{v} \cdot \mathbf{v}^{\sf T}.
\end{equation*}
Furthermore, for any vector $\mathbf{x} \in \mathbb{R}^3$,
\begin{gather*}
  \big(\mathbf{v} \cdot \mathbf{v}^{\sf T}\big) \cdot \mathbf{x} = \mathbf{v} \big(\mathbf{v}^{\sf T} \cdot \mathbf{x}\big)  = (\mathbf{v} \cdot \mathbf{x}) \mathbf{v}, \\
  \begin{bmatrix} \mathbf{v} \end{bmatrix}_\times^2 \cdot \mathbf{x} = \mathbf{v} \times (\mathbf{v}\times \mathbf{x}) = (\mathbf{v}\cdot \mathbf{x}) \mathbf{v} - \mathbf{x},
\end{gather*}
from which Eq.~\eqref{eq:dyadic_cross} immediately follows.
\end{proof}

\begin{lemma}
\label{lemma:cross}
Let $\mathbf{v}$ and $\mathbf{w}$ be two arbitrary unit vectors such that $(\mathbf{v} \cdot \mathbf{w})^2 \neq 1$, and $\mathbf{n} \equiv \mathbf{v} \times \mathbf{w}$. Then,
\begin{multline}
  \label{eq:lemcross}
  \widehat{\mathbf{n}} \otimes \widehat{\mathbf{n}} = \mathcal{I} - \frac{1}{1+\mathbf{v} \cdot \mathbf{w}} \Big \{  \mathbf{v}\otimes \mathbf{w} +\mathbf{w}\otimes \mathbf{v} \\ + 2 \big(\widehat{\mathbf{v-w}}\big) \otimes \big(\widehat{\mathbf{v-w}}\big) \Big \},
\end{multline}
where
\begin{equation*}
  \widehat{\mathbf{x}} \equiv \frac{\mathbf{x}}{\lVert \mathbf{x} \rVert} \qquad \forall \mathbf{x} \in \mathbb{R}^3,
\end{equation*}
with $\lVert \cdot \rVert$ the Euclidean norm.
\end{lemma}

\begin{proof}
Lemma~\ref{lemma:dyadic} immediately yields
\begin{equation}
  \label{eq:nxn1}
   \widehat{\mathbf{n}} \otimes \widehat{\mathbf{n}} = \mathcal{I} + \frac{1}{1-(\mathbf{v} \cdot \mathbf{w})^2} \begin{bmatrix} \mathbf{n} \end{bmatrix}_\times^2,
\end{equation}
where we used
\begin{equation*}
   \widehat{\mathbf{n}} = \frac{\mathbf{n}}{\sqrt{1-(\mathbf{v} \cdot \mathbf{w})^2}}.
\end{equation*}
Furthermore, for any vector $\mathbf{x} \in \mathbb{R}^3$,
\begin{align*}
  \begin{bmatrix} \mathbf{n} \end{bmatrix}_\times^2 \cdot \mathbf{x}
  &= (\mathbf{v} \times \mathbf{w}) \times \big\{ (\mathbf{v} \times \mathbf{w}) \times \mathbf{x} \big\} \\
  &= (\mathbf{v} \times \mathbf{w}) \times \big\{(\mathbf{v}\cdot \mathbf{x}) \mathbf{w} - ( \mathbf{w}\cdot\mathbf{x}) \mathbf{v}\big\}\\ 
  &= \big\{ (\mathbf{v}\cdot \mathbf{w}) (\mathbf{w}\cdot \mathbf{x}) - (\mathbf{v}\cdot \mathbf{x})\big\} \mathbf{v} \\
  & \qquad\qquad\qquad +\big\{ (\mathbf{v}\cdot \mathbf{w}) (\mathbf{v}\cdot \mathbf{x}) - (\mathbf{w}\cdot \mathbf{x})\big\} \mathbf{w} \\
  &= (\mathbf{v}\cdot \mathbf{w}) (\mathbf{v} \otimes \mathbf{w} + \mathbf{w} \otimes \mathbf{v}) \cdot \mathbf{x} \\
  &\qquad\qquad\qquad - (\mathbf{v}\otimes\mathbf{v}+\mathbf{w}\otimes\mathbf{w})\cdot \mathbf{x}.
\end{align*}

Thus,
\begin{align}
  \begin{bmatrix} \mathbf{n} \end{bmatrix}_\times^2 
  &=  (\mathbf{v}\cdot \mathbf{w}) (\mathbf{v} \otimes \mathbf{w} + \mathbf{w} \otimes \mathbf{v})-(\mathbf{v}\otimes\mathbf{v}+\mathbf{w}\otimes\mathbf{w}) \nonumber \\
  &= (\mathbf{v}\cdot \mathbf{w}-1)  (\mathbf{v} \otimes \mathbf{w} + \mathbf{w} \otimes \mathbf{v}) \nonumber \\
  \label{eq:n_x2}
  &\qquad\qquad\qquad - (\mathbf{v}-\mathbf{w})\otimes(\mathbf{v}-\mathbf{w}).
\end{align}
Plugging Eq.~\eqref{eq:n_x2} into Eq.~\eqref{eq:nxn1} leads to
\begin{multline}
  \label{eq:nxn2}
  \widehat{\mathbf{n}} \otimes \widehat{\mathbf{n}} = \mathcal{I} - \frac{1}{1+\mathbf{v} \cdot \mathbf{w}}(\mathbf{v} \otimes \mathbf{w} + \mathbf{w} \otimes \mathbf{v}) \\
 - \frac{1}{1-(\mathbf{v} \cdot \mathbf{w})^2}(\mathbf{v}-\mathbf{w})\otimes(\mathbf{v}-\mathbf{w}),
\end{multline}
and substituting $\lVert \mathbf{v}-\mathbf{w}\rVert^2 = 2\,(1-\mathbf{v} \cdot \mathbf{w})$ into Eq.~\eqref{eq:nxn2} directly yields Eq.~\eqref{eq:lemcross}.
\end{proof}

\begin{theorem}
\label{theorem:qi}
Let $\mathcal{N}_k$ be the normal tensor associated with the $i$-th pair of consecutive inter-monomer bonds,
\begin{equation*}
  \mathcal{N}_k \equiv \Big(\widehat{\mathbf{t}_k\times \mathbf{t}_{k+1}}\Big) \otimes \Big(\widehat{\mathbf{t}_k\times \mathbf{t}_{k+1}}\Big),
\end{equation*}
and $\mathbf{u}_k \equiv \mathbf{t}_{k+1}-\mathbf{t}_k$. Then,
\begin{equation}
  \label{eq:qi_exp}
  \mathcal{N}_k = \mathcal{I} - \frac{\mathbf{t}_k\otimes \mathbf{t}_k + \mathbf{t}_{k+1}\otimes \mathbf{t}_{k+1}}{2} - \widehat{\mathbf{u}}_k \otimes  \widehat{\mathbf{u}}_k  + \mathcal{O}\big(\mathbf{u}_k^2\big).
\end{equation}
\end{theorem}

\begin{proof}
Using Lemma~\ref{lemma:cross},
\begin{equation}
  \label{eq:qi_full}
  \mathcal{N}_k = \mathcal{I} - \frac{1}{1+\mathbf{t}_k \cdot \mathbf{t}_{k+1}}\big(\mathbf{t}_k \otimes \mathbf{t}_{k+1} + \mathbf{t}_{k+1} \otimes \mathbf{t}_k + 2\widehat{\mathbf{u}}_k \otimes  \widehat{\mathbf{u}}_k\big). 
\end{equation}
It is straightforward to check that
\begin{equation*}
  \mathbf{t}_k \otimes \mathbf{t}_{k+1} + \mathbf{t}_{k+1} \otimes \mathbf{t}_k = \mathbf{t}_k \otimes \mathbf{t}_k + \mathbf{t}_{k+1} \otimes \mathbf{t}_{k+1} + \mathcal{O}\big(\mathbf{u}_k^2\big),
\end{equation*}
along with 
\begin{equation*}
  \mathbf{t}_k \cdot \mathbf{t}_{k+1} = 1+ \mathcal{O}\big(\mathbf{u}_k^2\big).
\end{equation*}
The Taylor expansion of Eq.~\eqref{eq:qi_full} to first order in $\mathbf{u}_k$ thus immediately yields Eq.~\eqref{eq:qi_exp}.
\end{proof}

\begin{corollary}
\label{corollary:contour}
Let $l_b$ be the fixed separation distance between two consecutive monomers along the chain and $\langle \cdot \rangle_c$ the (discrete) contour average,
\begin{equation*}
  \langle \cdot \rangle_c \equiv \frac{1}{N_m -2} \sum_{k=1}^{N_m-2} \; \cdot \;,
\end{equation*}
where $N_m$ is the total number of monomers per chain. Then, in the limit of long polymers ($N_m \gg 1$),
\begin{equation}
  \label{eq:q_contour}
  \big\langle \mathcal{N}_k \big\rangle_c = \mathcal{I} -\big \langle \mathbf{t}_k \otimes \mathbf{t}_k \big \rangle_c - \big\langle \widehat{\mathbf{u}}_k \otimes \widehat{\mathbf{u}}_k  \big\rangle_c + \mathcal{O}\big(l_b^2C_m^2\big),
\end{equation}
with $C_m \equiv \sqrt{\big\langle \mathbf{u}_k^2 \big\rangle_c}/l_b$ the root mean square curvature of the chains.
\end{corollary}

\begin{proof}
The Kremer-Grest (KG) definition of the discrete local curvature $C_k$ reads as~\cite{grest1986molecular,kremer1990dynamics}
\begin{equation}
  \label{eq:curv}
  C_k \equiv \frac{\Vert \mathbf{t}_{k+1}-\mathbf{t}_{k} \rVert}{l_b} = \frac{\sqrt{\mathbf{u}_k^2}}{l_b}.
\end{equation}
Eq.~\eqref{eq:q_contour} then trivially follows from Eqs.~\eqref{eq:qi_exp} and~\eqref{eq:curv}, using $C_m = \sqrt{\big\langle C^2_k\big\rangle_c}$ and
\begin{equation*}
  \big\langle \mathbf{t}_k\otimes \mathbf{t}_k + \mathbf{t}_{k+1}\otimes \mathbf{t}_{k+1} \big\rangle_c = 2 \big\langle\mathbf{t}_k\otimes \mathbf{t}_k \big\rangle_c +\mathcal{O} \bigg ( \frac{1}{N_m} \bigg),
\end{equation*}
where the extremal terms of order $1/N_m$ may be neglected in the long-chain limit.
\end{proof}

\begin{theorem}
\label{theorem:qi_unconfined}
Let $\langle \cdot \rangle$ be the canonical thermodynamic average,
\begin{equation*}
  \langle \cdot \rangle \equiv \langle \langle \cdot \rangle_c \rangle_{\rho, T}
\end{equation*}
with $\langle \cdot \rangle_{\rho, T}$ the ensemble average over the polymer configurational space at fixed temperature $T$ and density $\rho$. In the limit of long, unconfined chains ($R \gg l_b$, $\rho \to 0$), 
\begin{equation}
  \label{eq:q_uca}
  \big\langle \mathcal{N}_k\big\rangle = \frac{I-\big\langle \mathbf{t}_k \otimes \mathbf{t}_k \big\rangle}{2} + \mathcal{O}\bigg(\frac{l_b}{l_p}\bigg),
\end{equation}
with $l_p$ the polymer persistence length.
\end{theorem}

\begin{proof}
Let $\Theta_k$ be the angle between two consecutive bonds $\mathbf{t}_k$ and $\mathbf{t}_{k+1}$,
\begin{equation*}
  \cos\Theta_k \equiv \mathbf{t}_k \cdot \mathbf{t}_{k+1}.
\end{equation*}
Neglecting any spontaneous chain curvature potentially induced by the confining membrane ($R \gg l_b$), it follows from the local definition of the persistence length~\cite{hsu2010standard} that, for sufficiently stiff filaments ($l_p \gg l_b$),
\begin{equation}
  \label{eq:theta_trans}
  \big\langle\cos\Theta_k\big\rangle_{\rho, T}  \xrightarrow[\rho \to 0]{}  \exp \bigg\{ - \frac{l_b}{l_p(T)} \bigg \} = 1 + \mathcal{O} \bigg( \frac{l_b}{l_p} \bigg).
\end{equation}
Note that in the case of finite densities within the nematic stability range, deflections of the chain by the surrounding polymers typically induce a further inhibition of transverse fluctuations~\cite{egorov2016insight}, so that Eq.~\eqref{eq:theta_trans} may generally provide an upper bound for the variations of $\Theta_k$. While similar considerations have been suggested to potentially favor the appearance of hairpin defects~\cite{milchev2018nematic}, we find no evidence of hairpin formation in any of the systems investigated here, and hence neglect the probability of their occurrence in the following discussion. Thus, 
\begin{equation}
  \label{eq:curvature_ave}
  \big\langle C^2_m\big\rangle_{\rho, T} = \frac{\big\langle\big\langle\mathbf{u}_k^2\big\rangle_c\big\rangle_{\rho, T}}{l_b^2} = \frac{2\big\langle 1-\cos\Theta_k\big\rangle}{l_b^2} = \mathcal{O}\bigg(\frac{1}{l_b l_p}\bigg).
\end{equation}
Furthermore, let $\mathbf{t}_{\perp k}$ be an arbitrary unit vector such that $\mathbf{t}_k \cdot \mathbf{t}_{\perp k} = 0$. We may rewrite $\mathbf{u}_k$ in the form
\begin{equation}
  \label{eq:u_proj}
   \mathbf{u}_{k} = u_{\perp k} \, \mathbf{t}_{\perp k} + u_{\independent  k} \, \big(\mathbf{t}_k \times \mathbf{t}_{\perp k}\big) + \mathcal{O}\bigg( \frac{l_b}{l_p}\bigg),
\end{equation}
where we used
\begin{equation*}
     \big\langle\mathbf{u}_{k}\cdot\mathbf{t}_k \big\rangle_{\rho, T} = \big\langle1-\mathbf{t}_{k}\cdot\mathbf{t}_{k+1} \big\rangle_{\rho, T} =  \mathcal{O}\bigg( \frac{l_b}{l_p} \bigg).
\end{equation*}
The KG bending energy penalty per chain then reads as
\begin{equation}
  \label{eq:kg_bend}
  U^{\rm bend}_{\rm poly} = \frac{\kappa_b}{2} \sum_{k=1}^{N_m-2} \mathbf{u}_k^2 =  \frac{\kappa_b}{2} \sum_{k=1}^{N_m-2} \big( u_{\perp k}^2 + u_{\independent k}^2\big) + \mathcal{O}(k_BT),
\end{equation}
with $\kappa_b \equiv  k_B T l_p/ l_b$ the polymer flexural modulus. Assimilating the transverse fluctuation components $u_{\perp k}$ and $u_{\independent k}$ to decoupled degrees of freedom, we obtain
\begin{gather} 
  \label{eq:u_ave}
  \big \langle u_{\perp k} u_{\independent k} \big \rangle_{\rho, T} = \big \langle u_{\perp k}\big \rangle_{\rho, T} \big \langle u_{\independent k}\big \rangle_{\rho, T} = 0,\\
    \label{eq:u2_ave}
  \big \langle u_{\perp k}^2\big \rangle_{\rho, T} = \big \langle u_{\independent k}^2\big \rangle_{\rho, T} = \frac{l_b}{l_p},
\end{gather}
where Eq.~\eqref{eq:u2_ave} results from the equipartition theorem. Note that Eq.~\eqref{eq:u_ave} assumes that the chains bear no local curvature at rest, consistently with Eq.~\eqref{eq:theta_trans}, while Eq.~\eqref{eq:u2_ave} further neglects additional Hamiltonian contributions beyond Eq.~\eqref{eq:kg_bend} which may arise from potential polymer-polymer and polymer-membrane interactions. Hence, Eqs.~\eqref{eq:u_ave} and~\eqref{eq:u2_ave} are only expected to hold in the so-called \textit{unconfined chain} regime~\cite{tortora2018incorporating}, in which local polymer conformations are largely unaffected by the presence of surrounding chains or membrane walls. In this case, Eqs.~\eqref{eq:u_proj},~\eqref{eq:u_ave} and~\eqref{eq:u2_ave} lead to
\begin{align}
  \big \langle \widehat{\mathbf{u}}_k \otimes \widehat{\mathbf{u}}_k \big\rangle_{\rho, T} &=  \frac{1}{2} \big \langle \big(\mathbf{t}_k \times \mathbf{t}_{\perp k}\big) \otimes \big(\mathbf{t}_k \times \mathbf{t}_{\perp k}\big) \big\rangle_{\rho, T} \nonumber \\
  &  \qquad\qquad\qquad +  \frac{1}{2} \big \langle \mathbf{t}_{\perp k} \otimes \mathbf{t}_{\perp k} \big\rangle_{\rho, T} +  \mathcal{O}\bigg( \frac{l_b}{l_p}\bigg) \nonumber \\
   \label{eq:uxu_ave}
  &= \frac{\mathcal{I} - \big\langle \mathbf{t}_k \otimes \mathbf{t}_k \big\rangle_{\rho, T}}{2} + \mathcal{O}\bigg( \frac{l_b}{l_p}\bigg),
\end{align}
where we used Lemma~\ref{lemma:cross}. Plugging Eqs.~\eqref{eq:curvature_ave} and~\eqref{eq:uxu_ave} into Eq.~\eqref{eq:q_contour} finally yields Eq.~\eqref{eq:q_uca}.
\end{proof}

\begin{corollary}
\label{corollary:qi_eig}
Let $\mathcal{Q}_k$ be the standard nematic OP tensor,
\begin{equation*}
  \mathcal{Q}_k \equiv \frac{3 \big(\mathbf{t}_k \otimes \mathbf{t}_k\big) - \mathcal{I}}{2}.
\end{equation*}
Then, for long polymers ($N_m \gg 1$) in the unconfined chain approximation (UCA),
\begin{equation}
  \label{eq:q_rel}
  \big\langle \mathcal{Q}_k\big\rangle  = \mathcal{I} - 3\big\langle \mathcal{N}_k\big\rangle + \mathcal{O}\bigg(\frac{l_b}{l_p}\bigg).
\end{equation}
Hence, the tensors $\mathcal{Q} \equiv \big\langle \mathcal{Q}_k\big\rangle$ and $\mathcal{N}\equiv\big\langle \mathcal{N}_k\big\rangle$ generally share the same eigenvectors, with their respective ascending eigenvalues $\big\{\lambda_{ i}\big\}$  and $\big\{\mu_{ i}\big\}$ related through
\begin{equation}
\label{eq:eig_perp}
\begin{gathered}
    \lambda_{ 1} = 1-3\,\mu_3, \\
    \lambda_{ 2} = 1-3\,\mu_2, \\
    \lambda_{ 3} = 1-3\,\mu_1.
\end{gathered}
\end{equation}
\end{corollary}

\begin{proof}
Eqs.~\eqref{eq:q_rel}-\eqref{eq:eig_perp} follow directly from Eq.~\eqref{eq:q_uca}. 
\end{proof}
Note that while Theorem~\ref{theorem:qi} and Corollary~\ref{corollary:contour} may be derived based solely on geometrical considerations, and are therefore quite generally valid for $N_m \gg 1$, the additional thermodynamic assumptions underlying Theorem~\ref{theorem:qi_unconfined} and Corollary~\ref{corollary:qi_eig} restrict their applicability to phases in which the UCA may reasonably hold --- i.e., in which local chain fluctuations do not significantly deviate from those expected from the equipartition theorem in the dilute regime (Eq.~\eqref{eq:u2_ave}). However, it is shown in Figs.~\ref{figS2}d-f that Eq.~\eqref{eq:eig_perp} is remarkably well satisfied for all systems considered, with relative discrepancies of less than $5\,\%$ in the corresponding eigenvalues being observed only for the longest chains studied ($l_c = 64\,\sigma$, Fig.~\ref{figS2}f). This observation further evidences the suitability of the UCA in our case, which provides the basis of the Fynewever-Yethiraj density functional theory employed here~\cite{tortora2018incorporating}.

%----------------------------------------------------------------------------------------
%	REFERENCE LIST
%----------------------------------------------------------------------------------------


%----------------------------------------------------------------------------------------